\documentclass[journal]{IEEEtran}
\usepackage{cite}

\ifCLASSINFOpdf
\else
\fi
\usepackage{amsmath}
\usepackage{algorithmic}

\usepackage{array}

\usepackage{caption}
\usepackage{subcaption}
\usepackage{graphicx}
\usepackage{amsmath}

\usepackage{amsmath,amssymb,amsfonts}
\usepackage{array}
\usepackage{enumitem}

\graphicspath{{eps/}{figures/}}

\usepackage[export]{adjustbox}
\usepackage{tikz}
\usepackage{pgfplots}
\usepackage{amsmath,amssymb,amsfonts}
\usepackage{mathrsfs} 
\usepackage{mathtools}
\usepackage{commath}
\usepackage{hyperref}
\usepackage{caption}
\usepackage{float}
\usepackage[linesnumbered]{algorithm2e}
\usepackage{hyperref}
\usepackage{balance}

\usepackage{multirow}
\usepackage{amsthm}
\newtheorem{theorem}{Theorem}
\newtheorem{lemma}{Lemma}

\newtheorem{remark}{Remark}

\newtheorem{problem}{Problem}
\newtheorem{proposition}{Proposition}
\newtheorem{definition}{Definition}
\newtheorem{property}{Property}
\usetikzlibrary {arrows.meta}

\begin{document}
%
\title{Distributed Coverage Control of Constrained Constant-Speed Unicycle Multi-Agent Systems}
%
%
%
\author{
    Qingchen Liu,
    Zengjie Zhang$\textsuperscript{*}$,
    Nhan Khanh Le,
    Jiahu Qin,
    Fangzhou Liu,
    Sandra Hirche
    \thanks{\textsuperscript{*}Corresponding author.}
    \thanks{This work was supported in part by the National Natural Science Foundation of China (No. 62373123) and the Ningbo Technology Project (No. 2023Z042).}
	\thanks{Q. Liu and J. Qin are with the Department of Automation, University of Science and Technology of China, Hefei, China. Email:{\tt\footnotesize \{qingchen\_liu, jhqin\}@ustc.edu.cn}.}
	\thanks{Z. Zhang is with the Department of Electrical Engineering, Eindhoven University of Technology, Netherlands. Email: {\tt\footnotesize z.zhang3@tue.nl}.}
	\thanks{N. Le and S. Hirche are with the Chair of Information-Oriented Control, Technical University of Munich, Munich, Germany. Email: {\tt\footnotesize \{nhankhanh.le, hirche\}@tum.de}.}
	\thanks{F. Liu is with Ningbo Institute of Intelligent Equipment Technology Company Ltd, Ningbo, China. Email: {\tt\footnotesize fangzhouliu.ac@gmail.com}.}

	}
\maketitle{}

\begin{abstract}
This paper proposes a novel distributed coverage controller for a multi-agent system with constant-speed unicycle robots (CSUR). The work is motivated by the limitation of the conventional method that does not ensure the satisfaction of hard state- and input-dependent constraints and leads to feasibility issues for multi-CSUR systems. In this paper, we solve these problems by designing a novel coverage cost function and a saturated gradient-search-based control law. Theoretical proofs are provided to guarantee that the CSURs ultimately move to the optimal coverage configuration without moving out of the covered domain. The controller is implemented in a distributed manner based on a novel communication standard among the agents. A series of simulation studies are conducted to validate the correctness of our theory by showing the efficacy of the proposed coverage controller in different initial conditions and with various control parameters. A comparison study in simulation reveals the advantage of the proposed method over the conventional method in terms of avoiding infeasibility. The experimental study verifies the applicability of the method to real robots. The development procedure of the method from theoretical analysis to experimental validation provides a novel framework for multi-agent system coordinate control with complex dynamics.
\end{abstract}



%
\IEEEpeerreviewmaketitle

\section{Introduction}\label{sec:intro}
%
%
%
%

\IEEEPARstart{T}{he} dynamic coverage of a target region using multiple robots is important for various practical applications such as event monitoring, production measuring, and resource allocation. The objective of coverage is to effectively allocate the robots in the region with a certain criterion optimized. Coverage has been achieved using the trajectory of a single robot~\cite{coombes2018optimal
}. However, multi-agent systems with networked robots are increasingly used due to higher efficiency and superior reliability. In multi-agent coverage, every spot of the target region is dominated by its closest agent. This renders a Centroidal Voronoi Tessellation, where each agent is positioned in the geometric center or the \textit{centroid} of a Voronoi partition~\cite{marx2015optimal}. Then, the multi-agent coverage problem can be solved by driving the agents to move along the negative gradient direction of the coverage criterion until they ultimately reach the optimal coverage configuration~\cite{cortes2004coverage}. 

Although gradient-based coverage control methods have been well developed for robots with simple dynamic models, such as quadcopters formulated as single integrators, or single-integrator robots (SIR), optimal coverage control using agents with complex dynamics remains an open and challenging problem~\cite{liu2017coverage}. From a control theory perspective, the coverage control of robots with complex dynamics is not trivial due to the modeling nonlinearity. The nonlinearity may deviate the motion of the robots from the specified negative gradient directions of the coverage cost function, which means that the robots may not ultimately reach the optimal coverage configuration, causing the failure of the coverage task.

In this paper, we are concerned with the constant-speed unicycle robots (CSURs), a representative type of robot that cruise at constant speeds~\cite{oh2015coordinated}, unlike SIRs which can freeze in fixed positions in the air. Thus, a CSUR is typically controlled to orbit around a fixed point~\cite {yuksek2016transition
}. In this case, optimal coverage can be realized by regulating each CSUR to orbit around the geometric center of its Voronoi partition~\cite{liu2017coverage}. Our focus on CSURs is motivated by the interest in optimal coverage using fixed-wing unmanned aerial vehicles (fUAV), a class of vehicles maneuvered by two fixed wings~\cite{qin2018circular}. Compared to a quadcopter, an fUAV can carry heavier loads and cruise faster with less power, offering higher efficiency, a longer air-borne time, and a larger coverage capability \cite{qin2018circular}. However, the conventional coverage controllers for SIRs do not apply to CSURs due to the possible infeasibility of the Centroidal Voronoi Tessellation when the orbiting centers of the CSURs move outside the target region before reaching the optimal coverage configuration. 

This infeasibility issue reflects the limitation of the conventional coverage control methods when applied to agents with complex dynamic models, such as CSURs. 
The main reason for the feasibility issue is that the orbiting movement of a CSUR renders an under-actuated dynamic model that brings up an additional state-dependent perturbation term. This term may deflect the desired movement direction of a CSUR and drive it outside the target region. This issue only appears in a multi-agent system with complex dynamics but not in one with simple and fully actuated agent dynamics like SIRs. To our best knowledge, the feasibility issue of a coverage control problem has not been well defined and studied by existing work, due to the lack of studies on the coverage control of complex agents. Fixing this requires a switching law, bringing discontinuity to the controller~\cite{liu2017coverage}. Another solution that has not been explored is to use several hard constraints to forcibly confine the orbiting centers of the CSURs within the target region. 
A barrier Lyapunov function (BLF)~\cite{zhao2013adaptive} is promising to incorporate these hard constraints into coverage control. This solution has not been studied in previous work due to the complexity of CSUR dynamics. 

Besides the feasibility issue, distributed realization is also important for coverage control. Practically, robots are not always fully connected, bringing challenges to centralized control approaches. A distributed controller that only requires local communication among adjacent agents is more robust to anomalies than centralized control. 
Although distributed coverage control for SIRs has been solved~\cite{schwager2008ladybug}, whether a multi-CSUR system admits a distributed coverage controller is still an open question. The main challenge lies in defining local communication among agents with complex dynamic models.

This paper solves these issues by proposing a novel distributed coverage controller for a multi-CSUR system. The work is based on solving a challenging mathematical problem: designing a coverage controller for a multi-robot system incorporating the requirements of nontrivial agent dynamic models, hard state-dependent constraints, saturated control inputs, and distributed implementation, which has not been solved in the literature. We have solved this problem by deriving a gradient-based optimal coverage controller with saturated inputs from a novel coverage cost. The coverage cost as a BLF, encodes the hard state-dependent constraints that guarantee the avoidance of infeasibility. We use control theory methods, namely the Lypapunov-based method and the controlled invariance theory~\cite{khalil2015nonlinear}, to rigorously prove that the controlled CSURs can reach the optimal configuration without moving outside the region. By overcoming this mathematical challenge, our work is expected to provide an effective baseline for the coverage control of generic robots with complicated dynamic models.

The rest of this paper is organized as follows. Sec.~\ref{sec:rw} addresses the challenge of the work based on reviewing the related work. Sec.~\ref{sec:pre} introduces the preliminaries and formulates the problem. Sec.~\ref{sec:main} proposes the theoretical results and Sec.~\ref{sec:sim} and Sec.~\ref{sec:exp} present the simulation and experimental studies. Finally, Section~\ref{sec:con} concludes the paper.

\textit{Notations}: $\mathbb{R}$ ($\mathbb{R}_+$ or $\mathbb{R}_{\geq 0}$) denotes the set of (positive or non-negative) real numbers. $\mathbb{N}$ ($\mathbb{N}^+$) denotes the set of (positive) natural numbers. For a real scalar $a \in \mathbb{R}$, $|a| \in \mathbb{R}_{\geq 0}$ is its absolute value. $x \in \mathbb{R}^n$ represents an $n$ dimensional vector and $A \in \mathbb{R}^{n \times m}$ is an $n$ by $m$ matrix. $\|x\|$ is the 2-norm of $x$ and $\|x\|_Q = \sqrt{x^{\!\top\!}Qx}$ is its weighted norm, $Q \in \mathbb{R}^{n \times n}$, $Q>0$. 
For a closed compact set $\Omega \in \mathbb{R}^n$, $\Omega$ represents the interior of $\Omega$ and $\partial \Omega$ is its boundary. For a set $\mathcal{A} \subset \Omega$, $\Omega - \mathcal{A}$ denotes the set difference of $\Omega$ and $\mathcal{A}$.


\section{Related Work}\label{sec:rw}
The optimal coverage problem is originally introduced in~\cite{cortes2004coverage} based on a facility location problem~\cite{chan1999facility} which also addresses the relation between its solution and a Centroidal Voronoi Tessellation. In~\cite{cortes2005coordination}, optimal coverage is defined as a coordination control problem for multi-agent systems with time-variant network topology and nonsmooth dynamics, based on which a general distributed coverage control law is proposed using nonsmooth gradient flows. 
Then, a general gradient searching law is designed for a team of SIRs~\cite{schwager2009optimal}. 
The gradient-based control framework is then extended to generic multi-agent coordination control problems in~\cite{schwager2009theory}. In~\cite{schwager2011unifying}, this control framework is further extended to various coverage cost criteria, with the non-convexity of the coverage problem highlighted. 

Later studies are dedicated to solving dynamic coverage control for nontrivial target domains or changeable environments.
In~\cite{arslan2016voronoi}, the coverage control problem is studied for a team of disk-shaped robots with heterogeneous sizes. 
In~\cite{mavrommati2017real}
, an adaptive controller is proposed for a time-variant coverage criterion. Besides, efforts are devoted to the optimal coverage over nontrivial geometric manifolds, like circles~\cite{song2020coverage}, 
spherical surfaces~\cite{li2013unified}, or arbitrary curves prescribed by vector-fields~\cite{zhou2014vector}. 
The work in~\cite{karatas2018optimal} attempts to seek a global optimal coverage solution. The work in \cite{arslan2019statistical} studies coverage control of robots with adjustable sensor ranges, leading to Voronoi partitions with soft margins instead of the conventional ones with clear boundaries.
In~\cite{atincc2020swarm}, a control scheme is proposed to ensure a smooth transference between coverage and other coordinate tasks. A survey on the previous development of multi-agent coverage control can be referred to in~\cite{wang2017coverage}. A review of optimal coverage control can be seen in~\cite{zhai2021multi}.

Recent work attempts to improve the flexibility of the control methods against imperfect environmental knowledge. 
In~\cite{abdulghafoor2021two}, a multi-level coverage controller is designed for an unknown density function, where a Gaussian mixture model is used for the approximation. A model-free coverage policy is solved using reinforcement learning in~\cite{hua2022drl}. 
Besides, additional constraints like timing costs and network connectivity are incorporated into optimal coverage~\cite{sun2022fixed}. 
Recent work also witnesses dynamic coverage control for multi-agent systems with complex dynamics, such as stratospheric airships~\cite{zhang2022region}. An observer-based coverage control method is presented in~\cite{sun2023observer} to improve the robustness of coverage against external disturbance. Besides, learning-based methods have been used for data-driven coverage control, such as reinforcement learning~\cite{aydemir2023multi} and the Gaussian process~\cite{xu2023multiagent}.

Compared to SIRs, the coverage control of complex agents attracts less attention. In~\cite{kwok2010unicycle, liu2017coverage}, coverage controllers are developed for CSURs, where the ultimate optimal coverage configuration corresponds to the solution where the orbiting centers of the CSURs coincide with the Voronoi centroids. The feasibility issue is solved using hard switching schemes which have obvious shortcomings. Firstly, they may lead to instability for an oddly shaped region due to the finite discrete-sampling rate. Secondly, they require a large control effort on the boundary of the target region, which is difficult to satisfy considering the practical control limits. Thirdly, the closed-loop system under hard switching is not robust to disturbances. 
To avoid hard switching in the controller inputs, a feasible solution is to formulate the feasibility requirement as a group of state-dependent constraints and encode them into the coverage controller using BLFs~\cite{wang2017safety}, which may result in a controller subject to the \textit{controlled invariance} property~\cite{blanchini1999set}. Although the barrier functions are widely applied to practical control systems due to the advantage of continuous control inputs, they have not been used for coverage control of complex agents. We believe that they can be used to solve the feasibility issue for multi-CSUR systems. Besides, the chattering attenuation technology used for sliding mode control can be used to generate smooth control inputs~\cite{zhang2022disturbance, najafabadi2023adaptive}.

\section{Preliminaries and Problem Formulation} \label{sec:pre}

This section introduces the mathematical preliminaries and formulates the problem to be studied.

\subsection{The Optimal Coverage Problem with Multiple Agents}

Let $\Omega \in \mathbb{R}^2$ be a closed convex polygonal set surrounded by $M \in \mathbb{N}^+$ linear edges, i.e.,
\begin{equation}\label{eq:lineqs}
\textstyle \Omega = \{ \omega \in \mathbb{R}^2 \left| h_j(\omega) \geq 0 \right.,\,\forall \, j \in \mathcal{M} \},
\end{equation}
where $\mathcal{M} \!=\! \{1,2,\cdots,M\}$ and $h_j(\omega) = b_j - a_j^{\!\top\!}\omega$,~$\omega \!\in\! \mathbb{R}^2$, $j \!\in\! \mathcal{M}$,
where $a_j \!\in\! \mathbb{R}^2$, $b_j \in \mathbb{R}$, are coefficients to denote the edges. Also, we denote the boundary and the interior of the region as $\partial \Omega =\{ \omega \in \mathbb{R}^2 | h_j(\omega) = 0, \,\exists \, j \in \mathcal{M} \}$ and $\mathrm{int} \,\Omega =\{ \omega \in \mathbb{R}^2 | h_j(\omega) > 0,\,\forall \, j \in \mathcal{M} \}$, respectively.
Note that $\mathrm{int} \,\Omega$ is open.
For simplicity, we assume the origin $O$ of the coordinate within $\Omega$ or on its boundary, i.e., $O \in \Omega$ without losing generality. Actually, for any other cases, we can always apply a coordinate transformation to make it satisfied for the new coordinate frame. Without losing generality, we prescribe $\|a_j\|=1$ and $b_j > 0$ for all edges $j \in \mathcal{M}$.
When $N\!\in\!\mathbb{N}^+$ agents are placed in the region $\Omega$ for coverage, the position of each is denoted as $z_k \in \mathbb{R}^2$, $k \in \mathcal{N}$, where $\mathcal{N} =\{1,2,\cdots, N\}$. We define $\mathcal{Z} =\{z_1,z_2,\cdots,z_N\}$, $z_i \neq z_j$ for any $i,j \in \mathcal{N}$, $i \neq j$, as a \textit{configuration} which is defined on a joint domain $\Omega^N$ $=$ $\underbrace{\Omega \times \cdots \times \Omega}_{N}$ with $\mathcal{Z} \in \Omega^N$ denoting $z_1 \in \Omega \cap z_2 \in \Omega \cap \cdots \cap z_N \in \Omega$. 

The objective of the optimal coverage problem is to properly locate the $N$ agents to minimize the following coverage cost,
\begin{equation}\label{eq:cost}
\textstyle	H(\mathcal{Z}) = \int_{\Omega} f(\omega, \mathcal{Z}) \Phi(\omega) \mathrm{d}\omega,~\mathcal{Z} \in \Omega^N,
\end{equation}
where $\omega \in \Omega$ denotes an event in the region $\Omega$, $\Phi: \Omega \rightarrow \mathbb{R}^+$ is a function that depicts the distribution of events $\omega \in \Omega$, and $f: \Omega \times \Omega^{N} \rightarrow \mathbb{R}^+$ is a function that assigns a real weight to an event $\omega \in \Omega$.  
In this paper, the weight function is~\cite{marx2015optimal},
\begin{equation}\label{eq:mfc}
\textstyle	f(\omega,\mathcal{Z}) = \min_{k\in \mathcal{N}} \frac{1}{2} \left\|\omega - z_k \right\|^2,~\mathcal{Z} \in \Omega^N,
\end{equation}
which calculates the squared Euclidean distance between an event $\omega \in \Omega$ and its closest agent. This is equivalent to splitting the region $\Omega$ into $N$ mutually exclusive Voronoi partitions $\Omega_1$, $\Omega_2$, $\cdots$, $\Omega_N$ using the $N$ agents. Each partition is defined as
\begin{equation}\label{eq:voronoi}
\textstyle	\Omega_{k} = \{\omega \in \Omega \left| \|\omega \!-\! z_k\|\! \leq\! \|\omega \!-\! z_i\|, \,\forall\, i \neq k, i \in \mathcal{N} \right. \}.
\end{equation}
Then, \eqref{eq:mfc} can be rewritten as $f(\omega,\mathcal{Z}) = \frac{1}{2} \left\|\omega - z_k \right\|^2$, $\mathrm{if} \, \omega \in \Omega_k$,
which takes off the minimum operator in~\eqref{eq:mfc} and converts it to a piece-wise quadratic form. Substituting this weight function $f(\omega,\mathcal{Z})$ into \eqref{eq:cost}, the coverage cost becomes
\begin{equation} \label{eqn:Hprimitive}
\textstyle		H(\mathcal{Z}) = \sum_{k = 1}^{N} \frac{1}{2} \int_{\Omega_k} \|\omega-z_k\|^2 \Phi(\omega) \mathrm{d}\omega
\end{equation}
which transfers the integration over the entire region $\Omega$ to the summation of the individual integrals on all Voronoi partitions. 
Note that Voronoi partitioning is usually performed using the geodesic distance measure which is only trivial for convex regions. It brings up additional complications for nonconvex regions~\cite{breitenmoser2010voronoi}. This paper only considers optimal coverage on a convex region $\Omega$ to avoid these complications.

Then, optimal coverage is achieved by placing the agents at the optimal configuration $\displaystyle \mathcal{Z}^* \!=\!\arg \min_{\mathcal{Z}\in \Omega^N} H(\mathcal{Z})$. Note that the cost function~\eqref{eqn:Hprimitive} is nonconvex and a global minimum solution is difficult to find~\cite{schwager2011unifying}. Similar to the previous work~\cite{schwager2009decentralized, liu2017coverage}, we use the following first-order optimization condition to solve the \textit{local optimal configuration} (LOC),
\begin{equation}\label{eq:problem}
\nabla H(\mathcal{Z}^*) = 0,
\end{equation}
where $\nabla H(\mathcal{Z}) \!=\! \left[\, \nabla_1 H(\mathcal{Z})~\cdots~\nabla_N H(\mathcal{Z}) \, \right]^{\!\top\!}$ is the gradient of the coverage cost with $\displaystyle \nabla_k H(\mathcal{Z})\!=\! \frac{\partial H(\mathcal{Z})}{\partial z_k}$ being its $k$-th element, $k \!\in\!\mathcal{N}$. Such solutions can be solved using gradient-based control laws~\cite{cortes2004coverage}. 
It is worth mentioning that there may exist multiple LOCs in domain $\Omega^N$. Also, its solution $\mathcal{Z}^*$ is not necessarily locally optimal but can be a saddle solution. Note that either finding the global optimum or inspecting the saddle solutions is a challenging topic beyond the scope of this paper. In this paper, we are concerned with the optimal coverage given by any LOC solution of~\eqref{eq:problem}.

\subsection{Distributed Coverage Controller for A Multi-SIR System}\label{eq:disc}

Given the Voronoi partitions defined in \eqref{eq:voronoi}, we say two partitions are adjacent if they share common boundaries, i.e., $\exists \, \omega \in \Omega$, $\omega \in \Omega_i \cap \Omega_j$. Based on this, we claim that agents $i,j\in\mathcal{N}$, $i \neq j$, are \textit{adjacent} if their Voronoi partitions $\Omega_i$ and $\Omega_j$ are adjacent. We define an adjacency mapping $\mathscr{A}:\mathcal{N} \rightarrow 2^{\mathcal{N}}$ to depict the adjacency relation between the agents. Specifically, $\mathscr{A}_k$, $k \in \mathcal{N}$ is the set of all adjacent agents of agent $k$. Note that the adjacency relation is bidirectional, i.e., for any $i,j \in \mathcal{N}$, $i\neq k$, $i \in \mathscr{A}_k \Leftrightarrow k \in \mathscr{A}_i$. Also, we define a commonly used set $\overline{\mathscr{A}_k} = \mathscr{A}_k \cup k$, $k \in \mathcal{N}$. The adjacency relation is needed to incorporate a common practical condition that communication can only be effective within a certain range~\cite{cortes2005coordination, schwager2009gradient}. For the optimal coverage problem, this range refers to the largest distance between adjacent agents, which renders a common and practical assumption that only adjacent agents can conduct bidirectional communication~\cite{schwager2009decentralized}.

Then, we discuss the solution to the optimal coverage problem~\eqref{eq:problem}.
According to~\cite{schwager2006distributed}, the $k$-th element of the gradient $\nabla H(\mathcal{Z})$ is calculated as
\begin{equation}\label{eq:gradient}
\nabla_k H(\mathcal{Z}) = M(\mathcal{Z}_{\!\overline{\mathscr{A}_k}}) (z_k \!-\! C(\mathcal{Z}_{\!\overline{\mathscr{A}_k}}) ),
\end{equation}
where $\mathcal{Z}_{\!\overline{\mathscr{A}_k}} \in \Omega^{|\overline{\mathscr{A}_k}|}$ is the set of all $z_j$ with $j \in \overline{\mathscr{A}_k}$ where $|\overline{\mathscr{A}_k}|$ is the number of elements in the finite set $\overline{\mathscr{A}_k}$, and $M(\mathcal{Z}_{\!\overline{\mathscr{A}_k}}) \in \mathbb{R}$ and $C(\mathcal{Z}_{\!\overline{\mathscr{A}_k}}) \in \mathbb{R}^2$ are the geometric mass and the centroid of the Voronoi partition $\Omega_k$, defined as $\displaystyle M(\mathcal{Z}_{\!\overline{\mathscr{A}_k}}) \!=\! \!\int_{\Omega_k}\!\!\!\Phi(\omega) \mathrm{d}\omega$ and $\displaystyle C(\mathcal{Z}_{\!\overline{\mathscr{A}_k}})  \!=\! \! \int_{\Omega_k}\! \frac{\omega\Phi(\omega) \mathrm{d}\omega}{M(\mathcal{Z}_{\!\overline{\mathscr{A}_k}})}$.
Here, we refer to $\mathcal{Z}_{\!\overline{\mathscr{A}_k}}$ as a \textit{partial configuration} since it only contains the positions of $z_k$ and its adjacent agents. It is noticed in \eqref{eq:gradient} that the computation of gradient $\nabla_k H(\mathcal{Z})$ only needs the positions of agent $k$ and its adjacent agents contained in $\mathcal{Z}_{\!\overline{\mathscr{A}_k}}$, which is an important property for the implementation of a distributed coverage controller to be discussed later. 

\begin{figure}[htbp]
\centering
\includegraphics[width=0.35\textwidth]{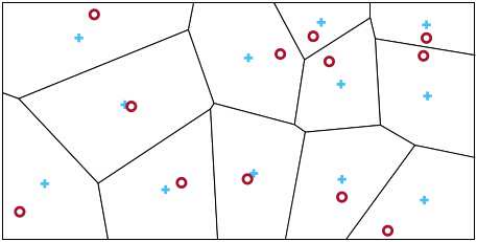}
\caption{Red `o' are agent positions and blue `+' are the Voronoi centroids. It does not illustrate a LOC since the `o' marks do not coincide with the blue `+' marks.}
\label{fig:voronoi}
\end{figure}

The relation among the agent positions, the Voronoi partition, and the centroids is illustrated in Fig.~\ref{fig:voronoi}.  Since $M(\mathcal{Z}_{\!\overline{\mathscr{A}_k}}) >0$ holds for all $k \in\mathcal{N}$, by solving $\nabla H(\mathcal{Z}) = 0$, we know that $\mathcal{Z}$ is a LOC if and only if
\begin{equation}\label{eq:equili}
z_k = C(\mathcal{Z}_{\!\overline{\mathscr{A}_k}} ),~\forall\,k \in \mathcal{N}.
\end{equation}
Therefore, if a configuration is a LOC, the agent positions and the Voronoi centroids must coincide.

A LOC indicated by condition~\eqref{eq:equili} can be found using the following gradient-based method
\begin{equation}\label{eq:closesir}
\dot{z}_k = - \nabla_k H(\mathcal{Z}),~k \in \mathcal{N},
\end{equation}
which is the main technical point of the conventional methods for the multi-agent coverage problem. For a multi-SIR system with the following single-integrator-based models~\cite{cortes2005coordination, schwager2009gradient},
\begin{equation}\label{eq:sin-int}
\dot{z}_k(t) = u_k(t),~k \in \mathcal{N},
\end{equation}
where $u_k(t)\in \mathbb{R}^2$ is the velocity of a SIR as its control input, a trivial optimal coverage controller can be designed as
\begin{equation}\label{eq:grad}
u_k(t) = - \nabla_k H(\mathcal{Z}),~k \in \mathcal{N}.
\end{equation}
It renders a distributed controller since the computation of $\nabla_k H(\mathcal{Z})$ only requires the measurements of agent $z_k$ and its adjacent agents.

\begin{remark}
The controllers of other multi-agent control problems, such as consensus~\cite{li2019survey}, formation~\cite{oh2015survey}, 
and distributed optimization~\cite{zhong2010asynchronous
}, can also be formulated as gradient-based forms~\cite{schwager2009optimal}. The unique challenge of coverage control compared to these problems is the nonconvex cost $H(\mathcal{Z})$.
\end{remark}

\subsection{The Dynamic Model of A CSUR}\label{eq:unipre} 
	
The dynamic model of a CSUR is described as~\cite{
bunjaku2017system},
\begin{equation}\label{eq:agent}
\begin{split}
\dot{\zeta}(t) &= v_0r(\theta) \\
\dot{\theta}(t) &= u(t),
\end{split}
\end{equation}
where $\zeta(t) \in \mathbb{R}^2$ and $\theta(t) \in \mathbb{R}$ are the position and the orientation of the CSUR at time $t \in \mathbb{R}_{\geq 0}$, respectively, $v_0\in \mathbb{R}^+$ is the constant linear speed of the robot, $u(t) \in \mathbb{R}$ is the angular velocity input of the robot, and $r(\theta) \!=\! [\,\cos(\theta)~\sin(\theta)\,]^{\!\top\!}$ is a transformation vector. It is easy to verify that $r(\theta)$ satisfies $\left\|r(\theta) \right\| = 1$ and $\displaystyle \frac{\partial^2 r(\theta)}{\partial \theta^2} = - r(\theta)$ for all $\theta \in \mathbb{R}$.

\begin{figure}[ht]
     \centering
     
\begin{tikzpicture}

\definecolor{shadowgray}{RGB}{105, 105, 105}

    \draw[->, >=Stealth, thick, color=shadowgray, densely dashed] (1-1.414/2, 1.414/2) arc (0:361:1);
    \draw[thick, color=red, densely dashed] (0, 0) arc (-45:-150:2);

    \node[circle,inner sep=0pt,minimum size=2mm,fill=black, very thick] (vct) at (-1.414/2, 1.414/2) {};
    
    \node[circle,inner sep=0pt,minimum size=2mm,draw=red, very thick] (robot) at (-1.414-1.7312, 1.414-1) {};
        \draw[<->, Stealth-{Rays[]}, densely dotted, very thick,color=black] (-1.414/2, 1.414/2) arc (-45:-150:1) node[pos=1,align=center, anchor=south]{$z(t)$};
    \draw[->,>=Stealth,color=red, very thick] (robot.center) -- node[pos=0,align=center, anchor=south east]{$\zeta(t)$} ([xshift=10,yshift=-10*1.7321] robot.center);
    \node[color=red, anchor=east]() at ([xshift=10,yshift=-10*1.7321] robot.center) {$\theta(t)$};
    \draw[<->, >=Stealth, thick, color=shadowgray] (-1.414/2, 1.414/2) -- node[pos=0.5,align=center, anchor=north]{$\displaystyle \frac{v_0}{|w_0|}$} (-1.414/2+0.9, 1.414/2+0.436);
    \draw[->, >=Stealth, thick, color=red, densely dashed] (-1.414, 1-1.57) arc (-90:-89:2);

    \node[color=shadowgray, anchor=south west]() at (-1.414/2+0.9, 1.414/2+0.436) {$u(t) \!\equiv\! \omega_0$};
\end{tikzpicture}
     \caption{The position $\zeta(t)$ (red `o') and the virtual center $z(t)$ (black `x') of a CSUR. The lines with arrows denote their trajectories. The arrow attached to $\zeta(t)$ denotes the robot's orientation $\theta(t)$. When $u(t) \!\equiv\! \omega_0$, the CSUR orbits along a time-invariant virtual center with a constant radius of $v_0/|\omega_0|$.}
     \label{fig:virtualCenter}
\end{figure}
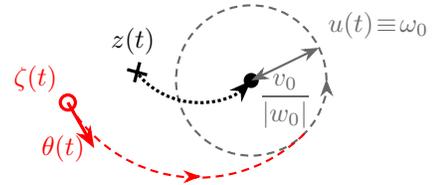

For the CSUR input $u(t)$ in (\ref{eq:agent}), we prescribe that $u(t)<0$ and $u(t)>0$ indicate clockwise and anticlockwise orientation directions, respectively. When $u(t) \equiv 0$, the CSUR moves along a straight line. Note that the robot model (\ref{eq:agent}) is under-actuated since the three-dimensional state $[\,\zeta^{\!\top\!}(t) ~\theta(t)\,]^{\!\top\!}$ is excited by a one-dimensional input signal $u(t)$. Also, it is impossible to let a CSUR freeze in a fixed position like a SIR since it always moves at a constant speed $v_0$. Following~\cite{seyboth2014collective, liu2017coverage}, we use the following virtual center of a CSUR, instead of its position $\zeta(t)$, to perform the coverage task,
\begin{equation}\label{eqn:virtualdy}
z(t) = \zeta(t) + \frac{v_0}{\omega_0} \frac{\partial r(\theta)}{\partial \theta} ,
\end{equation}
where $\omega_0 \in \mathbb{R}$, $\omega_0 \neq 0$ is a constant parameter that represents the nominal angular velocity of CSUR. Taking the derivative of (\ref{eqn:virtualdy}), the dynamic model of the virtual center is
\begin{equation} \label{eq:z_kDot}
\dot{z}(t) = \dot{\zeta}(t) + \frac{v_0}{\omega_0} \frac{\partial^2 r(\theta)}{\partial \theta^2} \dot{\theta}(t) = v_0r(\theta) - \frac{v_0}{\omega_0} r(\theta) u(t).
\end{equation}
The meaning of the virtual center $z(t)$ is not straightforward for an arbitrary robot trajectory $\zeta(t)$ but is clear for a special case $u(t) \equiv \omega_0$. Substituting it into \eqref{eq:z_kDot}, we have $\dot{z}(t) = 0$ which denotes that the virtual center $z(t)$ is a static point in this case. Then, equation \eqref{eqn:virtualdy} indicates that the robot is moving around $z(t)$ along a circular orbit with a linear speed $v_0$, an angular velocity $\omega_0$, and orbit radius $v_0/|\omega_0|$. Thus, $z(t)$ can be interpreted as the center of the circular orbit of the CSUR when it is a static point, which is why it is referred to as a \textit{virtual} center. The relation between the CSUR position $\zeta(t)$ and its virtual center $z(t)$ is illustrated in Fig.~\ref{fig:virtualCenter}.

Different from the CSUR position $\zeta(t)$ that has to always move at a constant linear speed, the virtual center $z(t)$ can remain static at a certain position when the CSUR is controlled with a constant input $u(t) \equiv \omega_0$, similar to the dynamics of a SIR. This implies the possibility of extending the existing results for SIRs to the virtual centers of CSURs. Therefore, in this paper, we refer to the CSUR virtual centers as \textit{CSUR agents} and use them for optimal coverage. Nevertheless, the dynamic model of a CSUR agent in \eqref{eq:z_kDot} is more complicated than that of a SIR in \eqref{eq:sin-int}, bringing up challenges to this extension. Sec.~\ref{sec:dccu} explains the challenges in detail.

\subsection{The Optimal Coverage Control of Multiple CSUR Agents}\label{sec:dccu}

Derived from \eqref{eq:z_kDot}, the dynamic model of each agent in a multi-CSUR system is described as
\begin{equation}\label{eq:virtdy}
\dot{z}_k(t) = v_k r(\theta_k) - \frac{v_k}{\omega_k} r(\theta_k) u_k(t),~k\in \mathcal{N},
\end{equation}
where $\theta_k(t), u_k(t) \!\in\! \mathbb{R}$ are the orientation and the control input of agent $k$, respectively, and $v_k \!\in\!\mathbb{R}^+$ and $\omega_k\!\in\!\mathbb{R}$ are predefined speed parameters. The nonlinear projection gain $r(\theta_k)$ and the additive perturbation term $v_k r(\theta_k)$ in~\eqref{eq:virtdy} complicate the coverage control problem, compared to SIRs. From (\ref{eq:agent}), $r(\theta_k)$ has a constant norm $1$, meaning that these nonlinear terms constantly perturb the agent velocity $\dot{z}(t)$ from the desired gradient-searching direction $-\nabla_k H(\mathcal{Z})$ and prevent $z_k(t)$ from converging to a LOC. Some $z_k(t)$ may even move out of $\Omega$, making Problem \eqref{eq:problem} infeasible. 
SIRs do not have such a feasibility issue due to their linear dynamic models in \eqref{eq:closesir}. As a result, the control law \eqref{eq:grad} always guides the SIRs to move inside the target region. Unlike SIRs, CSUR agents must be confined within the target region to guarantee feasibility. Moreover, their control inputs should satisfy saturation restrictions due to limited energy or resources. Based on this consideration, we formulate the following optimal coverage control problem for a multi-CSUR system.

\begin{problem}\label{pb:unictr}
Given a convex set $\Omega \subset \mathbb{R}^2$ defined in \eqref{eq:lineqs} and $N$ CSUR agents depicted by (\ref{eq:virtdy}), design a control law $u_k(t)$ for all $k \in \mathcal{N}$ to achieve the following objectives.
\begin{enumerate}
\item \label{it:pb:input} For all $k \in \mathcal{N}$ and $t\in \mathbb{R}_{\geq 0}$, the control inputs satisfy 
\begin{equation}\label{eq:inputsat}
\exists\,\overline{U}_k \in \mathbb{R}^+,~\mathrm{s.t.},~|u_k(t)| \leq \overline{U}_k,~\forall \,k\in \mathcal{N}.
\end{equation}
\item \label{it:pb:state} For all $t \in \mathbb{R}_{\geq 0}$, the agent configuration $\mathcal{Z}(t)$ satisfies
\begin{equation}\label{eq:inv}
\mathcal{Z}(t) \in \Omega^N,~\forall \mathcal{Z}(0) \in \Omega^N.
\end{equation}
\item \label{it:pb:equi} The agent configuration $\mathcal{Z}(t)$ asymptotically converges to a LOC $\mathcal{Z}^*$ specified by~\eqref{eq:problem}.
\end{enumerate}
\end{problem}

The main difference between Problem~\ref{pb:unictr} and the multi-SIR coverage problem in previous work~\cite{cortes2005coordination, schwager2009gradient} is reflected by the additional input- and state-dependent constraints specified in objectives \ref{it:pb:input}) and \ref{it:pb:state}). 
Additionally, the optimal coverage configuration $\mathcal{Z}^*$ is defined for the virtual centers of the CSURs, instead of their positions. When a LOC is achieved, the CSURs are expected to move along their circular orbits around their static virtual centers specified by the optimal configuration $\mathcal{Z}^*$. 
Problem~\ref{pb:unictr} is only concerned with a LOC instead of a globally optimal solution. Multiple LOC solutions may exist, possibly also including saddle solutions. Which LOC is ultimately reached and whether it is a saddle solution mainly depends on the initial conditions~\cite{du1999centroidal}. 

In this paper, we are only concerned with minimizing the coverage cost~\eqref{eqn:Hprimitive} without incorporating collision avoidance or time limits. These specifications render additional nontrivial challenges and are impractical to be fully addressed by this paper. A possible solution for collision avoidance may be shaping the coverage cost, for which artificial potential field~\cite{cao2022parameter} or control barrier functions~\cite{xiao2022sufficient} can be used. Besides, one can reduce the orbiting radius of the CSURs for a lower chance of collisions by heuristically selecting a small $v_k$ and a larger ${\omega_k}$ for $k\in \mathcal{N}$. Extensions to these challenging problems will be explored in future work.

\subsection{Positively Invariant Set and Tangent Cone}\label{sec:pist}

The \textit{positively invariant set} and the \textit{tangent cone} defined as follows are important to the analysis of the satisfaction of the hard state-dependent constraints specified by Problem~\ref{pb:unictr}.

\begin{definition}\label{eq:pis}
\cite{blanchini1999set}~$\mathcal{S} \subset \mathbb{R}^n$ is a positively invariant set for system $\dot{x}(t) = f(x(t))$ if $\forall \, x(0) \in \mathcal{S}$, $x(t) \subset \mathcal{S}$ for $t \in \mathbb{R}_+$.
\end{definition}

\begin{definition}
\cite{blanchini1999set}~The tangent cone of a convex set $\mathcal{S} \subset \mathbb{R}^n$ in $x \in \mathbb{R}^n$ is a set
\begin{equation}
\textstyle \mathscr{C}_{\mathcal{S}}(x) = \{ z \in \mathbb{R}^n \left| \lim_{\tau \rightarrow 0} \frac{1}{\tau}\mathscr{D}(x+\tau z, \mathcal{S}) = 0 \right. \},
\end{equation}
where $\mathscr{D}:\mathbb{R}^n \times 2^{\mathbb{R}^n} \rightarrow \mathbb{R}_{\geq 0}$ is a function that specifies the distance between a vector and a set,
\begin{equation}\label{eq:distf}
\textstyle \mathscr{D}(x,\mathcal{S}) = \mathrm{inf}_{s \in \mathcal{S}}\|x-s\|.
\end{equation}
\end{definition}

The hard state-dependent constraints in~\eqref{eq:inv} can be satisfied by forcing $\Omega$ to be a \textit{positively invariant set}. In this case, all tangent cones of $\Omega$ only contain safe moving directions of the CSUR agents, i.e., moving inside $\Omega$. Whether a closed set is positively invariant is determined by the following Lemma.

\begin{lemma}\label{lm:tc}
\cite{blanchini1999set}~Consider a system $\dot{x}(t) = f(x(t))$ of which each initial condition $x(0) \in \mathcal{X} \subseteq \mathbb{R}^n$ admits a globally unique solution. Then, a closed set $\mathcal{S} \subseteq \mathcal{X}$ is positively invariant for the system if and only if $f(x) \in \mathscr{C}_{\mathcal{S}}(x)$, $\forall \, x\in \partial \mathcal{S}$, where $\partial \mathcal{S}$ is the boundary of $\mathcal{S}$.
\end{lemma}

Lemma \ref{lm:tc} provides an easy approach to validate whether a designed controller achieves objective~\ref{it:pb:state}) of Problem~\ref{pb:unictr} by only investigating the tangent cone on the boundary of $\Omega$. Note that Lemma \ref{lm:tc} only applies to closed sets.

\section{Design of the Coverage Controller} \label{sec:main}

This section presents the main methods of this paper, including the off-LOC cost function, the novel coverage cost function, and the derived distributed coverage controller. 

\subsection{The Off-LOC Cost}\label{sec:off-loc}

We propose an off-LOC cost function which is important to derive the novel coverage cost function in the next subsection.
For any agent $k \in \mathcal{N}$ and its adjacent agents $\mathscr{A}_k$, the off-LOC cost function is defined as follows,
\begin{equation}\label{eq:ener}
\textstyle W(\mathcal{Z}_{\!\overline{\mathscr{A}_k}}) \!=\! \frac{1}{2} {\| z_k(t) \!-\! C(\mathcal{Z}_{\!\overline{\mathscr{A}_k}})\|_{Q_k}^2},\, \mathcal{Z}_{\!\overline{\mathscr{A}_k}} \!\in\! \Omega^{|\overline{\mathscr{A}_k}|},
\end{equation}
where $Q_k \!\in\! \mathbb{R}^{2\times 2}$ is a symmetrically positive-definite matrix. It can be verified that $W(\mathcal{Z}_{\!\overline{\mathscr{A}_k}}) \!\geq\! 0$, $\forall\, \mathcal{Z}_{\!\overline{\mathscr{A}_k}} \!\in\! \Omega^{|\overline{\mathscr{A}_k}|}$. Also, $W(\mathcal{Z}_{\!\overline{\mathscr{A}_k}}) \!=\! 0$ if and only if \eqref{eq:equili} is satisfied. 
For any agent $k \in \mathcal{N}$, the off-LOC cost function $W(\mathcal{Z}_{\!\overline{\mathscr{A}_k}})$ measures how far its virtual center $z_k$ is off its corresponding centroid $C(\mathcal{Z}_{\!\overline{\mathscr{A}_k}})$. It equals to zero only if $\mathcal{Z}_{\!\overline{\mathscr{A}_k}}$ belongs to a LOC. This is why it is named the \textit{off-LOC cost}.


\begin{proposition}\label{pr:w_0}
$W(\mathcal{Z}_{\!\overline{\mathscr{A}_k}})$ has the following properties for all $\mathcal{Z}_{\!\overline{\mathscr{A}_k}} \in \Omega^{|\overline{\mathscr{A}_k}|}$ and any $k \in \mathcal{N}$.

\noindent 1). There always exists $\overline{W} \in \mathbb{R}_+$, such that $W(\mathcal{Z}_{\!\overline{\mathscr{A}_k}}) < \overline{W}$.

\noindent 2). $W(\mathcal{Z}_{\!\overline{\mathscr{A}_k}}) > 0$ always holds if $z_k \in \partial \Omega$.

\noindent 3). There always exists $\epsilon \in \mathbb{R}_+$, $\displaystyle \epsilon < \min_{j \in \mathcal{M}} \sup_{\omega \in \Omega} h_j(\omega)$, such that $W(\mathcal{Z}_{\!\overline{\mathscr{A}_k}}) > 0$ holds for any $x_k \in \Omega - \Omega_{\epsilon}$,  where $\Omega_{\epsilon} \subset \Omega$ is a closed convex set defined as
\begin{equation}\label{eq:ome_ep}
\Omega_{\epsilon} = \{ \omega \in \mathbb{R}^2 \left| h_j(\omega) \geq \epsilon \right.,~\forall \, j \in \mathcal{M} \}.
\end{equation}
\end{proposition}

\begin{proof}
For property 1), we know any configuration defined in the region $\Omega$, i.e., $\mathcal{Z} \!\in\! \Omega^N$, corresponds to a certain Voronoi partition of $\Omega$, such that $\Omega_k \neq \varnothing$ and $M(\mathcal{Z}_{\!\overline{\mathscr{A}_k}}) > 0$ hold for all $k \in \mathcal{N}$. As a result, $z_k$ and $C(\mathcal{Z}_{\!\overline{\mathscr{A}_k}})$ are both bounded, meaning that $W(\mathcal{Z}_{\!\overline{\mathscr{A}_k}})$ always has an upper bound $\overline{W} \in \mathbb{R}_+$, $\forall \, k \in \mathcal{N}$. For property 2), we consider the inverse proposition, i.e., supposing that there exists $k \in \mathcal{N}$, such that $z_k \in \partial \Omega$, leading to $W(\mathcal{Z}_{\!\overline{\mathscr{A}_k}})=0$ or $z_k = C(\mathcal{Z}_{\!\overline{\mathscr{A}_k}})$. However, from the definition of $C(\mathcal{Z}_{\!\overline{\mathscr{A}_k}})$ in Sec.~\ref{eq:disc}, we know $C(\mathcal{Z}_{\!\overline{\mathscr{A}_k}}) \!\notin\! \partial \Omega$, violating the inverse proposition. Thus, the original proposition in 2) is satisfied. For 3), we know that $W(\mathcal{Z}_{\overline{\mathscr{A}_k}})$ is a continuous function of $z_k$ since $C(\mathcal{Z}_{\overline{\mathscr{A}_k}})$ is also continuous to $z_k$, according to \eqref{eq:ener}. Also, property 2) addresses that $W(\mathcal{Z}_{\!\overline{\mathscr{A}_k}}) > 0$ holds for any $z_k \in \partial \Omega$, $k \in \mathcal{N}$. Then, for any LOC $\mathcal{Z}^*=\{z_1^*, z_2^*, \cdots, z_N^*\}$, there always exists $\epsilon \!\in\! \mathbb{R}_+$, $\displaystyle \epsilon \!<\! \min_{j \in \mathcal{M}} \sup_{\omega \!\in\! \Omega} h_j(\omega)$, such that a). $\Omega_{\epsilon} \neq \varnothing$, and b). there exists $k \in \mathcal{N}$ such that $z_k \in \partial \Omega_{\epsilon}$ while $z_i \in \mathrm{int}\, \Omega_{\epsilon}$ for all $i \in \mathcal{N}$, $i\neq k$. For the smallest $\epsilon$ over all LOCs, we know that $W(\mathcal{Z}_{\!\overline{\mathscr{A}_k}})> 0$ holds for all $x_k \in \Omega - \Omega_{\epsilon}$.
\end{proof}

Proposition \ref{pr:w_0} provides several important statements on the off-LOC cost functions. Property 1) gives its upper limit as a positive constant $\overline{W}$. According to \eqref{eq:ener}, the value of $\overline{W}$ can be estimated by considering the extreme cases for all $k\in \mathcal{N}$ with $C(\mathcal{Z}_{\!\overline{\mathscr{A}_k}})$ being the geometric center of the target region and $z_k$ placed at the farthest convex of the target region to $C(\mathcal{Z}_{\!\overline{\mathscr{A}_k}})$. Property 2) indicates that LOC does not occur on the boundary $\partial \Omega$, and Property 3) ensures the existence of a margin $\Omega - \Omega_{\epsilon}$ around $\Omega$ where no LOC exists. They both address that all LOCs are inside $\Omega$ and do not appear in the marginal area close to its boundary. This is the foundation of our theoretical results in Sec.~\ref{sec:ana}.

Since $C(\mathcal{Z}_{\!\overline{\mathscr{A}_i}})$ is differentiable to $z_k$, $W(\mathcal{Z}_{\!\overline{\mathscr{A}_i}})$ is also differentiable to $z_k$, for $i, k \in \mathcal{N}$. According to \cite{du2006acceleration}, its partial derivative to $z_k$, i.e., $\displaystyle \nabla_k C(\mathcal{Z}_{\!\overline{\mathscr{A}_i}})\! =\! \frac{\partial C^{\!\top\!}(\mathcal{Z}_{\!\overline{\mathscr{A}_i}})}{\partial z_k}$ reads
\begin{equation}\label{eq:detC}
\nabla_k C(\mathcal{Z}_{\!\overline{\mathscr{A}_i}}) = \frac{D(\mathcal{Z}_{\overline{\mathscr{A}_i} }, z_k)}{M(\mathcal{Z}_{\!\overline{\mathscr{A}_i}})}  - P(\mathcal{Z}_{\overline{\mathscr{A}_i} }, z_k) C^{\!\top\!}(\mathcal{Z}_{\!\overline{\mathscr{A}_i}}), 
\end{equation}
where, for $z_i, z_k \in \Omega$, $i,k\in \mathcal{N}$, $i \neq k$, $z_i \neq z_k$,
\begin{subequations}\label{eq:dp}
\begin{equation}
D(\mathcal{Z}_{\overline{\mathscr{A}_i} }, z_k) = \int_{\partial \Omega_k^i } \frac{(\omega-z_k)\omega^{\!\top\!}}{\|z_k-z_i\|} \Phi(\omega) \mathrm{d} \omega,
\end{equation}
\begin{equation}
P(\mathcal{Z}_{\overline{\mathscr{A}_i} }, z_k) = \int_{\partial \Omega_k^i } \frac{\omega-z_k}{\|z_k-z_i\|} \Phi(\omega) \mathrm{d} \omega,
\end{equation}
\end{subequations}
where $\partial \Omega_k^i$ is the shared boundary of adjacent partitions $\Omega_i$, $\Omega_k$, $i,k \in \mathcal{N}$. 
Then, the gradient $\displaystyle \nabla_k W(\mathcal{Z}_{\!\overline{\mathscr{A}_i}})\!=\!\frac{\partial W\!(\mathcal{Z}_{\!\overline{\mathscr{A}_i}})}{\partial z_k}$ is
\begin{equation}\label{eq:W_par}
\small{
\nabla_k W(\mathcal{Z}_{\!\overline{\mathscr{A}_i}}) \!=\! \left\{ \!\begin{array}{ll}
\! ( I\! -\! \nabla_k C(\mathcal{Z}_{\!\overline{\mathscr{A}_i}})) {Q_k}(z_i\!-\!C(\mathcal{Z}_{\!\overline{\mathscr{A}_i}})), &\!\!\!\! i\!=\! k, \\
\displaystyle - \nabla_k C(\mathcal{Z}_{\!\overline{\mathscr{A}_i}}) {Q_k}(z_i\!-\!C(\mathcal{Z}_{\!\overline{\mathscr{A}_i}})), &\!\! i \!\neq\! k.
\end{array} \right.}
\end{equation}

\begin{proposition}\label{eq:distrib}
For any $i,k \in \mathcal{N}$, $i \neq k$, $\nabla_k C(\mathcal{Z}_{\!\overline{\mathscr{A}_i}})=0$ and $\nabla_k W(\mathcal{Z}_{\!\overline{\mathscr{A}_i}})=0$ hold, if $i \notin \mathscr{A}_k$ or $k \notin \mathscr{A}_i$.
\end{proposition}
\begin{proof}
According to \eqref{eq:dp}, for any $i,k \in \mathcal{N}$, $i \neq k$, we have $D(\mathcal{Z}_{\overline{\mathscr{A}_i} }, z_k) = 0$ and $P(\mathcal{Z}_{\overline{\mathscr{A}_i} }, z_k) = 0$ if $i \notin \mathscr{A}_k$ or $k \notin \mathscr{A}_i$. Substituting \eqref{eq:detC} into \eqref{eq:W_par}, we obtain $\nabla_k W(\mathcal{Z}_{\!\overline{\mathscr{A}_i}})=0$, proving this proposition.
\end{proof}

\begin{proposition}\label{pr:parboun}
$\| \nabla_k W(\mathcal{Z}_{\!\overline{\mathscr{A}_i}}) \|\!<\!\overline{W}_{\partial}$, $\exists \overline{W}_{\partial} \!\in\! \mathbb{R}, \forall k,i \in \mathcal{N}$.
\end{proposition}

\begin{proof}
In \eqref{eq:W_par}, it is noticed that $\nabla_k C(\mathcal{Z}_{\!\overline{\mathscr{A}_i}})$ is continuous and bounded since $M(\mathcal{Z}_{\!\overline{\mathscr{A}_i}}) > 0$ holds on $\Omega$ and $D(\mathcal{Z}_{\overline{\mathscr{A}_i} }, z_k)$, $P(\mathcal{Z}_{\overline{\mathscr{A}_i} }, z_k)$, and $C(\mathcal{Z}_{\!\overline{\mathscr{A}_i}})$ are all continuous and bounded. Thus, $\nabla_k W(\mathcal{Z}_{\!\overline{\mathscr{A}_i}})$ is also continuous and bounded. 
\end{proof}

\begin{remark}
Similar to Proposition \ref{pr:w_0}-1), the bounding scalar $\overline{W}_{\partial}$ can also be estimated by searching the extreme cases where $C(\mathcal{Z}_{\!\overline{\mathscr{A}_i}})$ being the geometric center of region $\Omega$ and $z_i$ placed at the region boundary $\partial \Omega$, for $i\!\in\! \mathcal{N}$.
\end{remark}


\subsection{The Coverage Cost}

This subsection proposes a novel coverage cost for coverage control of multiple CSURs. We give the mathematical form of the cost and introduce its important properties, followed by an intuitive interpretation of its underlying mechanism. The coverage cost is defined as the following BLF~\cite{zhao2013adaptive},
\begin{equation}  \label{eqn:V}
V(\mathcal{Z}) = \sum^{N}_{i=1}  \sum^{M}_{j=1} \frac{W(\mathcal{Z}_{\!\overline{\mathscr{A}_i}})}{h_j(z_i)},~\mathcal{Z} \in \mathrm{int}\, \Omega^N,
\end{equation}
where $W(\mathcal{Z}_{\!\overline{\mathscr{A}_i}})$ is an off-LOC cost function defined in Sec.~\ref{sec:off-loc} and $\mathrm{int}\, \Omega^N = \underbrace{\mathrm{int}\, \Omega \!\times\! \cdots \!\times\! \mathrm{int}\, \Omega}_{N}$ denotes the product of $N$ open sets. Note that $V(\mathcal{Z})$ is defined on an open domain and has the following properties. 

\begin{property}\label{pp:Vp}
The coverage cost function $V(\mathcal{Z})$ satisfies the following conditions for all $\mathcal{Z} \in \mathrm{int} \, \Omega^N$.

\noindent 1). $V(\mathcal{Z}) = 0$ holds if and only if $\mathcal{Z}$ is a LOC that satisfies the condition in~\eqref{eq:equili}, otherwise $V(\mathcal{Z})>0$.

\noindent 2). For any $\overline{V} \in \mathbb{R}_+$, there always exists $\epsilon \in \mathbb{R}_+$, such that $V(\mathcal{Z}) > \overline{V}$ holds for any $h_j(z_i) < \epsilon$, $i \in \mathcal{N}$, $\exists \, j \in \mathcal{M}$.

\noindent 3). For any $\epsilon \in \mathbb{R}+$, $\displaystyle \epsilon < \min_{j \in \mathcal{M}} \sup_{\omega \in \Omega} h_j(\omega)$, there always exists $V_{\epsilon} \in \mathbb{R}_+$, such that $V(\mathcal{Z}) < V_{\epsilon}$ holds for all $\mathcal{Z} \in \mathrm{int} \,\Omega_{\epsilon}^N$, where $\Omega_{\epsilon}$ is the closed set defined in \eqref{eq:ome_ep}.

\end{property}

The proof for Property \ref{pp:Vp} is not provided considering the straightforward boundedness of the off-LOC cost functions $W(\mathcal{Z}_{\!\overline{\mathscr{A}_k}})$, $k\in \mathcal{N}$, addressed in Proposition \ref{pr:w_0}. The property \ref{pp:Vp}-1) indicates the equivalence between $V(\mathcal{Z})=0$ and $\mathcal{Z}$ being a LOC. Properties \ref{pp:Vp}-2) and \ref{pp:Vp}-3) address that the coverage cost $V(\mathcal{Z})$ becomes unbounded if and only if an element of $\mathcal{Z}$ approaches the region boundary $\partial \Omega$. For the property \ref{pp:Vp}-3), calculating the upper bound $V_{\epsilon}$ is very challenging since it is not only dependent on the configuration of all agents $\mathcal{Z}$ but also related to the value of $\epsilon$. Nevertheless, it can be approximated by random sampling on the region $\Omega_{\epsilon}$.

The novel coverage cost in \eqref{eqn:V} is defined as a BLF~\cite{tee2009barrier} that 
decays to zero when the virtual centers $\mathcal{Z}_{\overline{\mathscr{A}_i}}$ for all $i\in \mathcal{N}$ reach a LOC. It approaches infinity when any virtual center $z_i$ gets close to any boundary $j \in \mathcal{M}$ of $\Omega$ since $h_j(z_i)$ approaches zero. This property is critical for designing a feasible coverage controller for multiple CSUR agents.

\subsection{The Coverage Controller}\label{eq:ccsisc}

Given the novel coverage cost defined in \eqref{eqn:V}, We design the following controller for the optimal coverage control problem \ref{pb:unictr} with the multi-CSUR system \eqref{eq:virtdy},
\begin{equation}
\label{eq:u_k}
\begin{split}
u_k(t) = \omega_k + \gamma_k \omega_k \, \rho ( \sigma(\mathcal{Z}, \theta_k) |\delta_k ),
\end{split}
\end{equation}
where $\sigma(\mathcal{Z}, \theta_k) \!=\! r^{\!\top}\!(\theta_k) \nabla_k V(\mathcal{Z})$, $\gamma_k \in \mathbb{R}^+$ is the control gain, $\delta_k \in \mathbb{R}^+$ is a boundary layer scalar, $\rho:\mathbb{R} \rightarrow (-1,1)$ is the following Sigmoid function commonly used for smooth control with saturation constraints~\cite{zhang2022disturbance}, 
\begin{equation}\label{eq:sigmoid}
\rho(x|\delta) = \frac{x}{|x|+\delta},~x \in \mathbb{R},~\delta \in \mathbb{R}^+,
\end{equation}
and $\displaystyle \nabla_k V(\mathcal{Z})\!=\! \frac{\partial V(\mathcal{Z})}{\partial z_k}$, for $\mathcal{Z} \in \mathrm{int}\, \Omega^N$, is the $k$-th element of the coverage cost gradient $\nabla V(\mathcal{Z})$, $k \in \mathcal{N}$, calculated as 
\begin{equation}\label{eq:parV0}
\nabla_k V(\mathcal{Z}) \!=\! \sum^{M}_{j=1} \left(\sum^{N}_{i=1} \frac{\nabla_k W(\mathcal{Z}_{\!\overline{\mathscr{A}_i}})}{h_j(z_i)} \!+\! a_j \frac{W(\mathcal{Z}_{\!\overline{\mathscr{A}_k}}) }{h_j^2(z_k)} \right).
\end{equation}

Compared to the conventional multi-SIR coverage controller in Eq.~\eqref{eq:sin-int}, the proposed CSUR controller in Eq.~\eqref{eq:u_k} is also gradient-based. Nevertheless, it has a different control gain $-\gamma_k \omega_k r^{\!\top\!}(\theta_k)/ (\left| r^{\!\top\!}(\theta_k)  \nabla_k V(\mathcal{Z}) \right| \!+\! \delta ) $ and an addition $w_k$. These terms are used to correct the deviation of CSURs from the desired negative gradient directions. The usage of a Sigmoid function ensures the smoothness of the control inputs. Compared to other constraint functions, such as the hyperbolic tangent function commonly used in machine learning, a sigmoid function is easier to implement. Besides, the following property ensures saturated control inputs.

\begin{property}\label{pp:inputsat}
The control input $u_k(t)$, in (\ref{eq:u_k}), for all $k \in \mathcal{N}$, is bounded by $|u_k(t)-\omega_k| < \gamma_k\omega_k$, for all $t \in \mathbb{R}_{\geq 0}$.
\end{property}

Property \ref{pp:inputsat} is straightforward to verify considering the continuity of $\rho(\cdot|\delta)$ on $\mathbb{R}$ and the property $\left|\rho(x|\delta)\right| < 1$ for any $x \in \mathbb{R}$ with any parameter $\delta \in \mathbb{R}^+$. It indicates that the proposed controller (\ref{eq:u_k}) is subject to the input-dependent constraint $|u_k(t)-\omega_k| < \gamma_k\omega_k$ which leads to $|u_k(t)| < \left(1+\gamma_k\right)\omega_k$. To ensure the input saturation constraint (\ref{eq:inputsat}), we may as well specify
\begin{equation}
(1+\gamma_k)\,\omega_k \leq \overline{U}_k.
\end{equation}
We adjust the control gain $\gamma_k$ or the nominal angular velocity $\omega_k$ for all $k\!\in\! \mathcal{N}$ to achieve objective \ref{it:pb:input}) of Problem \ref{pb:unictr}.

The proposed controller \eqref{eq:u_k} can keep the CSUR agents at a LOC. Substituting \eqref{eq:W_par} into (\ref{eq:parV0}), we obtain 
\begin{equation}\label{eq:partialVz}
\begin{split}
\nabla_k V(\mathcal{Z}) = \, & \sum_{j=1}^M \left( \frac{{Q_k}(z_k - C(\mathcal{Z}_{\!\overline{\mathscr{A}_k}}))}{h_j(z_k)} + \frac{a_jW(\mathcal{Z}_{\!\overline{\mathscr{A}_k}})}{h^2_j(z_k)} \right. \\
&-\sum_{i=1}^N  \left. \nabla_k C(\mathcal{Z}_{\!\overline{\mathscr{A}_i}}) \frac{ {Q_k}(z_i-C(\mathcal{Z}_{\!\overline{\mathscr{A}_i}}))}{h_j(z_i)} 
\right).
\end{split}
\end{equation}

Note that the gradient $\nabla V(\mathcal{Z})$ is continuous, considering the continuity of the linear constraint functions $h_j(z_i)$, the virtual centers $z_i(t)$, and the Voronoi centroids $C(\mathcal{Z}_{\!\overline{\mathscr{A}_i}})$, $i \in \mathcal{N}$, $j \in \mathcal{M}$. Also, $\nabla V(\mathcal{Z})$ satisfies the following condition.
\begin{proposition}\label{pp:vz0}
For any $\mathcal{Z} \in\mathrm{int}\, \Omega^N$, $\nabla V(\mathcal{Z})\! =\! 0$ holds if and only if \eqref{eq:equili} holds.
\end{proposition}
\begin{proof}
The sufficiency of this proposition is straightforward to verify by substituting \eqref{eq:equili} into (\ref{eq:partialVz}). For the necessity, we investigate (\ref{eq:parV0}). Since all CSURs have identical dynamic models, the number $N$ should not affect the equality of (\ref{eq:parV0}). Therefore, according to \eqref{eq:partialVz}, considering $h_j(z_i) > 0$ for all $z_i \in \Omega$, $i \in \mathcal{N}$, and all $j \in \mathcal{M}$, we can infer that $\nabla V(\mathcal{Z}) =0$ holds if and only if $W(\mathcal{Z}_{\overline{\mathscr{A}_k}}) =0$ and $z_k = C(\mathcal{Z}_{\overline{\mathscr{A}_k}})$ hold for all $k \in \mathcal{N}$, which is equivalent to \eqref{eq:equili}. This verifies the necessity of Proposition \ref{pp:vz0}.
\end{proof}

Considering a condition where $\nabla V(\mathcal{Z})=0$ which leads to $u_k(t)=\omega_k$ for all $k\in \mathcal{N}$, we know that all CSURs are orbiting around fixed virtual points as explained in Sec.~\ref{eq:unipre}. Meanwhile, Proposition~\ref{pp:vz0} indicates that $\nabla V(\mathcal{Z})=0$ holds if and only if CSURs are located in a LOC. Therefore, the proposed controller~\eqref{eq:u_k} can keep CSURs in a LOC. In Sec.~\ref{sec:ana}, we will prove that the controller~\eqref{eq:u_k} can drive the CSUR agents to a LOC from any initial positions. 


\section{Analysis of the Coverage Controller}\label{sec:ana}

In this section, we use control theory methods to rigorously prove that the proposed coverage controller in Eq.~\eqref{eq:u_k} can lead the CSURs to an LOC without them driving out of the covered region. Intuitive explanations will be given to address their underlying mechanisms. Finally, we interpret the distributed implementation of the proposed controller.

\subsection{Guarantee of State-Dependent Constraints}\label{sec:invri}

In this subsection, we address that the proposed controller in \eqref{eq:u_k} avoids the infeasibility issue, i.e., the CSURs never move out of the covered region during the formation of the optimal coverage, using the invariance theory~\cite{khalil2015nonlinear}, a common concept in control theory to describe the subjection to state-dependent constraints. The readers are suggested to refer to \cite{blanchini1999set} for details. Here, we directly give the mathematical results.

Substituting the controller (\ref{eq:u_k}) into (\ref{eq:virtdy}), the closed-loop dynamic model of each CSUR agent is
\begin{equation} \label{eqn:z_k}
\dot{z}_{k}(t)  = - \gamma_k \omega_k\, r(\theta_k)\, \rho (\sigma(\mathcal{Z}, \theta_k)|\delta_k ),~k \in \mathcal{N}.
\end{equation}
We use the invariance property introduced in Lemma~\ref{lm:tc} to validate whether the closed-loop dynamic model \eqref{eqn:z_k} achieves objective \ref{it:pb:state}) in Problem \ref{pb:unictr}. However, Lemma~\ref{lm:tc} only applies to closed sets but all CSUR agents are defined in an open domain $\mathrm{int}\, \Omega$. This brings up the challenges of the invariance analysis. In this paper, we perform an indirect manner by investigating the invariance of a closed subset $\Omega_{\epsilon}$ defined in \eqref{eq:ome_ep} with a small scalar $\epsilon$, rendering the following theorem.

\begin{theorem}\label{th:inv}
There always exists $\epsilon_0 \in \mathbb{R}_+$, such that for all $\epsilon < \epsilon_0$, 
$\Omega_{\epsilon} \neq \varnothing$ and $\Omega_{\epsilon}$ is positively invariant for system~\eqref{eqn:z_k}.
\end{theorem}

\begin{proof}
The critical point is to solve the tangent cone $\mathscr{C}_{\Omega_{\epsilon}}(z_k)$ for any $z_k \in \Omega_{\epsilon}$, $k \in \mathcal{N}$, with given $\epsilon$ and validate whether the trajectory admitted by \eqref{eqn:z_k} falls in $\mathscr{C}_{\Omega_{\epsilon}}(z_k)$. Inspired by Lemma~\ref{lm:tc}, we just need to calculate $\mathscr{C}_{\Omega_{\epsilon}}(z_k)$ for $z_k \in \partial \Omega_{\epsilon}$ since $\mathscr{C}_{\Omega_{\epsilon}}(z_k) = \mathbb{R}^2$ for all $z_k \in \mathrm{int} \, \Omega_{\epsilon}$. Without losing the generality, we assume that $z_k$ is closest to the boundary $\partial \Omega$ among all agent positions $z_r$, $r \in \mathcal{N}$, i.e., we always assign $\epsilon$ such that $z_k \in \partial \Omega_{\epsilon}$ while $z_r \in \Omega_{\epsilon}$, $\forall \, r \in \mathcal{N}$, $r \neq k$. 

Proposition~\ref{pr:w_0} implies the existence of a $\epsilon_0 \!\in\! \mathbb{R}_+$, such that $W(\mathcal{Z}_{\overline{\mathscr{A}_k}}) \!>\! 0$ for all $\epsilon \!<\! \epsilon_0$, if $z_k \!\in\! \partial \Omega_{\epsilon}$. Also, $\Omega_{\epsilon} \neq \varnothing$ is ensured if $\displaystyle \epsilon_0 < \min_{j \in \mathcal{M}} \sup_{\omega \in \Omega} h_j(\omega)$. Thus, we define the following function for $z_k \in \partial \Omega_{\epsilon}$, $\epsilon < \epsilon_0$ with an arbitrary vector $\iota \in \mathbb{R}^2$,
\begin{equation}\label{eq:scrV}
\mathscr{V}_{\epsilon}(z_k, \iota) = \frac{\overline{h}^2\!(z_k)}{W_k} \iota^{\!\top\!}\nabla_k V(\mathcal{Z}),
\end{equation}
where $W_k$ is the brief form of $W(\mathcal{Z}_{\overline{\mathscr{A}_k}})$, $k \in \mathcal{N}$, and $\displaystyle \overline{h}(z_k) = \min_{j \in \mathcal{M}} h_j(z_k)$.
Substituting~\eqref{eq:parV0} into \eqref{eq:scrV}, we have
\begin{equation*}
\mathscr{V}_{\epsilon}(z_k, \iota)
\!=\! \sum^{M}_{j=1} \! \left(\sum_{i \in \overline{\mathscr{A}_k}}  \frac{\overline{h}^2\!(z_k)}{h_j(z_i)}\frac{\iota^{\!\top\!}\nabla_k W_i}{W_k}\!+\! \iota^{\!\top\!}a_j \frac{\overline{h}^2\!(z_k)}{h_j^2(z_k)} \right).
\end{equation*}
According to Propositions~\ref{pr:w_0} and~\ref{pr:parboun}, we know that both $\nabla_k W_i$, $\forall \, i \in \overline{\mathscr{A}_k}$, and $W_k$ are all bounded for $k\in \mathcal{N}$. 
Thus, we know that $\mathscr{V}_{\epsilon}(z_k, \iota)$ has the following limit as $\epsilon \rightarrow 0$,
\begin{equation}
\mathscr{V}(z_k, \iota) =\lim_{\epsilon \rightarrow 0} \mathscr{V}_{\epsilon}(z_k, \iota) = \iota^{\!\top\!} a_r,
\end{equation}
where $\displaystyle r\!=\!\arg \min_{j \in \mathcal{M}} h_j(z_k)$ indexes the edge to which $z_k$ is the most close. Be reminded that $a_r$ is the normal vector of not only the $r$-th edge of $\Omega$ but also the $r$-th edge of $\Omega_{\epsilon}$ for all $\epsilon < \epsilon_0$. Moreover, the direction of $a_r$ points inside $\Omega$ and $\Omega_{\epsilon}$.
When $\iota \!=\! \dot{z}_k \in \mathbb{R}^2$, we have $\mathscr{V}(z_k, \dot{z}_k) \!=\! \dot{z}_k^{\!\top\!} a_r$ which is the inner product of the system trajectory direction $\dot{z}_k$ and the normal vector $a_j$. The sign of $\mathscr{V}(z_k, \dot{z}_k)$ indicates whether $\dot{z}_k$ points inside $\Omega$ and $\Omega_{\epsilon}$ for $\epsilon< \epsilon_0$. 
Then, we obtain the following relation between $\mathscr{V}(z_k, \dot{z}_k)$ and the distance function $\mathscr{D}$ in \eqref{eq:distf} used to define the tangent cone $\mathscr{C}_{\Omega_{\epsilon}}(z_k)$, 
\begin{equation*}
\textstyle \lim_{\tau \rightarrow 0} \frac{1}{\tau} \mathscr{D}(z_k\!+\! \tau \dot{z}_k, \Omega_{\epsilon}) > 0 \Leftrightarrow \mathscr{V}(z_k, \dot{z}_k) > 0,
\end{equation*}
\begin{equation*}
\textstyle \lim_{\tau \rightarrow 0} \frac{1}{\tau} \mathscr{D}(z_k\!+\! \tau \dot{z}_k, \Omega_{\epsilon}) = 0 \Leftrightarrow \mathscr{V}(z_k, \dot{z}_k) \leq 0,
\end{equation*}
for any $z_k \in \partial \Omega_{\epsilon}$ with any $\epsilon < \epsilon_0$ and $\dot{z}_k \in \mathbb{R}^2$. 
This indicates that the tangent cone $\mathscr{C}_{\Omega_k}(z_k)$ for any $z_k \!\in\! \partial \Omega_{\epsilon}$ and $\epsilon < \epsilon_0$ is $\mathscr{C}_{\Omega_{\epsilon}}(z_k) \!=\! \{ \dot{z}_k \!\in\! \mathbb{R}^2 | \mathscr{V}(z_k,\dot{z}_k) \!\leq\! 0 \}$.

Now, let us validate whether the trajectory direction $\dot{z}_k$ admitted by~\eqref{eqn:z_k} falls in the tangent cone $\mathscr{C}_{\Omega_k}(z_k)$. Substituting the closed-loop dynamics \eqref{eqn:z_k} into \eqref{eq:scrV}, we have
\begin{equation}\label{eq:clos_ve}
\mathscr{V}_{\epsilon}(z_k, \dot{z}_k) =
- \frac{\displaystyle \gamma_k \omega_k \overline{h}^2(z_k) |\sigma(\mathcal{Z}, \theta_k) |}{\displaystyle W_k (1 + \delta_k /|\sigma(\mathcal{Z}, \theta_k) | ) }.
\end{equation}
Note that $\displaystyle \lim_{\epsilon \rightarrow 0} \!\frac{\overline{h}^2\!(z_k) \sigma(\mathcal{Z}, \theta_k)}{W_k} \!=\! \lim_{\epsilon \rightarrow 0}\! \mathscr{V}_{\epsilon}(z_k, r(\theta_k)) \!=\! r^{\!\top\!}\!(\theta_k)a_r$
and $\displaystyle \lim_{\epsilon \rightarrow 0} \frac{1}{|\sigma(\mathcal{Z}, \theta_k) |}  \!=\! 0$. Taking the limit of \eqref{eq:clos_ve}, we have
\begin{equation}
\lim_{\epsilon \rightarrow 0} \mathscr{V}_{\epsilon}(z_k, \dot{z}_k) = \mathscr{V}(z_k, \dot{z}_k) = -\gamma_k \omega_k \left|r^{\!\top\!}(\theta_k)a_r \right| \leq 0,
\end{equation}
which implies that the dynamic model~\eqref{eqn:z_k} ensures
\begin{equation}\label{eq:invcond}
\dot{z}_k \in \mathscr{C}_{\Omega_{\epsilon}}(z_k), z_k \in \partial \Omega_{\epsilon},~\forall \, \epsilon < \epsilon_0.
\end{equation}
According to Lemma~\ref{lm:tc}, the condition \eqref{eq:invcond} means that $\Omega_{\epsilon}$ is invariant for $z_k$, i.e., for any initial condition $z_k(0) \in \Omega_{\epsilon}$, $z_k(t) \in \Omega_{\epsilon}$ holds for all $t \in \mathbb{R}_+$. Note that this generally holds for any agent $k \in \mathcal{N}$ closest to the boundary $\partial \Omega$. Thus, we prove that $\Omega_{\epsilon}$ is positively invariant for system~\eqref{eqn:z_k}.
\end{proof}
	
\begin{figure}[ht]
     \centering
     \usetikzlibrary {arrows.meta}

\resizebox{!}{3.2cm}{
\begin{tikzpicture}

\definecolor{shadowgray}{RGB}{105, 105, 105}
\definecolor{shadowblue}{RGB}{229, 242, 255}
\definecolor{shadowred}{RGB}{255, 229, 229}
\definecolor{vsgray}{RGB}{242, 242, 242}
\definecolor{vsgrayplus}{RGB}{220, 220, 220}

    \coordinate(linetip1) at (-2.94, 1.18);
    \coordinate(linetip2) at (3.06, -0.82);
    
    \path[fill=vsgray] (linetip1) -- ++ (-0.5, -1) --++ (0, -3) -- ++(5.7, 0) -- (linetip2) --cycle;
    
    \path[fill=vsgrayplus] (linetip1) -- (-3, 1) -- (3, -1) -- (linetip2) --cycle;
    
    \draw[gray, very thick] (linetip1) -- node[pos=0.2, anchor=south west, black]{$\partial \Omega$} (linetip2); %
    \draw[gray, densely dotted, very thick] (-3, 1) -- (3, -1);

    \node[thick, minimum height=0.3cm, minimum width=0.5cm, draw=shadowgray, fill=vsgrayplus] (l1) at (1, 1) {};
    \draw[color=shadowgray, thick, densely dotted] (l1.south west) -- (l1.north east);
    \path[fill=vsgray] (l1.south west) -- (l1.south east) -- (l1.north east) --cycle;
    \node[thick, minimum height=0.3cm, minimum width=0.5cm, draw=shadowgray] (l1) at (1, 1) {};
    
    \node[thick, minimum height=0.3cm, minimum width=0.5cm, draw=shadowgray, fill=vsgray] (l2) at (1, 0.5) {};
    
    \node[anchor=west] () at (l1.east) {$\Omega$};
    \node[anchor=west] () at (l2.east) {$\Omega_{\epsilon}$};
    
    \coordinate (robot1) at (-1.6, -1.2);
    \draw[very thick, dotted, color=shadowgray, fill=shadowblue] (robot1) circle (1);
    \foreach \x in {0, ..., 11}
        \draw[->, >=Stealth, dashed, color=shadowgray] (robot1) -- ++ (18.43+30*\x:1);
    \node[circle,inner sep=0pt,minimum size=2mm,draw,fill=blue] (r1) at (robot1) {};
    \node[anchor=north east, blue]() at (r1.center){$\dot{z}_1$};
    \draw [->, >=Stealth, very thick, blue] plot [smooth, tension=1] coordinates { (-2, -2.5) (robot1) (-2.5, 0.3)};
    \node[anchor=north east, blue]() at (-2.5, 0.3){$\dot{z}_1$};
    
    \coordinate (robot2) at (1.2, -0.4);
    \draw[very thick, dotted, color=shadowgray, fill=shadowred] (1.2-0.9163-0.03, -0.4+0.4-0.09) arc (-198.4376:-18.4376:1);
    \foreach \x in {0, ..., 5}
        \draw[->, >=Stealth, dashed, color=shadowgray] (robot2) -- ++ (-18.43-15-30*\x:1);
    \node[circle,inner sep=0pt,minimum size=2mm,draw,fill=red] (r2) at (robot2) {};
    \node[anchor=south west, red]() at (r2.center){$z_2$};
    \draw [->, >=Stealth, very thick, red] plot [smooth, tension=1] coordinates { (2, -2) (robot2) (0.5, -1.5)};
    \node[anchor=north east, red]() at (0.5, -1.5){$\dot{z}_2$};

\end{tikzpicture}
}
     \caption{An example of $\Omega_{\epsilon}$ as a positive invariant set. For any $z_1, z_2 \!\in\! \Omega_{\epsilon}$, their moving directions $\dot{z}_1, \dot{z}_2$ (the solid arrows) are confined in their corresponding tangent cones (the sector areas). The dashed arrows in the tangent cones indicate the allowed moving directions. The tangent cone of any interior state like $z_1$ is $\mathbb{R}^2$, allowing arbitrary moving directions. However, that of a marginal state on the boundary of $\Omega_{\epsilon}$ like $z_2$ only allows moving inside $\Omega_{\epsilon}$. }
     \label{fig:pinv}
\end{figure}
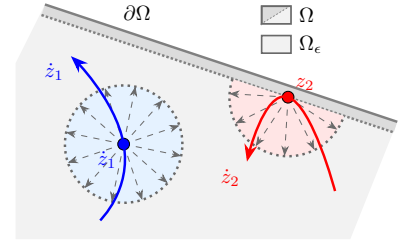

From the perspective of control theory, Theorem~\ref{th:inv} implies the proposed controller~\eqref{eq:u_k} ensures that there always exists a positively invariant subset of $\Omega$, forming the foundation of satisfying the state-dependent constraint in \eqref{eq:inv}. From a practical perspective, this theorem indicates that the CSURs always stay inside the covered region and away from the boundary during the entire period, as long as they initiate inside the region, reflecting the \textit{invariance} of the feasibility of the coverage over time.
The invariance property proved by this theorem means that both objectives \ref{it:pb:input}) and \ref{it:pb:state}) of Problem~\ref{pb:unictr} are achieved by the proposed controller \eqref{eq:u_k}. 
Fig.~\ref{fig:pinv} shows an example of a positively invariant set and how it confines the motion of the CSUR agents.

\subsection{Convergence to Optimal Coverage} \label{sec:stability}

In Sec.~\ref{sec:dccu}, we have addressed that the optimal coverage is achieved when the CSUR agents reach a LOC. The following theorem implies that optimal coverage can be ultimately achieved if the agents start inside the covered region.


\begin{theorem}\label{th:stab}
For CSUR agents defined in (\ref{eq:virtdy}) with control laws as (\ref{eq:u_k}), a LOC defined in (\ref{eq:equili}) is an asymptotically stable equilibrium.
\end{theorem}

\begin{proof}
We take the time derivative of the coverage cost $V(\mathcal{Z})$ defined in (\ref{eqn:V}) as follows,
\begin{equation}\label{eq:vdot}
\dot V(\mathcal{Z}) =  \sum^{N}_{k=1} \dot{z}_k^{\!\top\!} \nabla_k V(\mathcal{Z}).
\end{equation}
Substituting (\ref{eqn:z_k}) into (\ref{eq:vdot}), we have
\begin{equation}\label{eq:v_dot_nega}
\begin{split}
\dot V(\mathcal{Z}) 
&=- \sum^{N}_{k=1} \gamma_k \omega_k \rho ( \sigma(\mathcal{Z}, \theta_k) | \delta_k) \sigma(\mathcal{Z}, \theta_k)\\
&=- \sum^{N}_{k=1}\frac{ \gamma_k \omega_k | \sigma(\mathcal{Z}, \theta_k) |^2 }{ |\sigma(\mathcal{Z}, \theta_k) | +\delta_k} \leq 0.
\end{split}
\end{equation}
We notice that $\dot{V}(\mathcal{Z}) = 0$ holds if and only if 
\begin{equation}\label{eq:rV}
\sigma(\mathcal{Z}, \theta_k) = r^{\!\top\!}\!(\theta) \nabla_k V(\mathcal{Z}) = 0,~\forall \, k \in \mathcal{N},
\end{equation}
for which either $\displaystyle \nabla_k V(\mathcal{Z}) = 0$ or $\displaystyle \nabla_k V(\mathcal{Z}) \neq 0$ but $r(\theta)$ and $\displaystyle \nabla_k V(\mathcal{Z})$ are orthogonal. 
We take the time derivative of both sides of (\ref{eq:rV}) and obtain
\begin{equation}\label{eq:devVV}
\dot{r}^{\!\top\!}(\theta) \nabla_k V(\mathcal{Z}) + r^{\!\top\!}(\theta) \frac{\partial \nabla_k V(\mathcal{Z})}{\partial z^{\!\top\!}_k} \dot{z}_k = 0,~k \in \mathcal{N}.
\end{equation}
Note that any trajectory of the multi-CSUR system starting from a configuration satisfying both \eqref{eq:rV} and \eqref{eq:devVV} remains in this configuration, indicating this configuration as a positively invariant set. Substituting \eqref{eqn:z_k} into \eqref{eq:devVV}, we find that the conditions \eqref{eq:rV} and \eqref{eq:devVV} both hold if and only if $\nabla V(\mathcal{Z})\!=\!0$ or $\mathcal{Z}$ is a LOC, according to Proposition
\ref{pp:vz0}. This implies that the set $\mathcal{L}$ that unites all LOCs is a positively invariant set, according to Definition \ref{eq:pis}. Besides, since only a LOC can satisfy both conditions \eqref{eq:rV} and \eqref{eq:devVV}, $\mathcal{L}$ is the largest positively invariant set in $\Omega^N$.

Since property 2) of Proposition \ref{pr:w_0} indicates that no LOCs are positioned on the boundary of the region $\Omega$, we may as well consider a compact set $\Omega_{\epsilon}$ as defined in \eqref{eq:ome_ep} with any sufficient small $\epsilon$, within which $\mathcal{L}$ is still the largest positively invariant set. Also, note that $\dot{V}(\mathcal{Z}) \leq 0$ always holds on the compact set $\Omega_{\epsilon}$, according to \eqref{eq:v_dot_nega}. Therefore, according to the La Salle invariant principle~\cite{khalil2015nonlinear}, we know that any trajectory of the system that starts in $\Omega_{\epsilon}^N$ will ultimately approach some LOC in set $\mathcal{L}$. This implies that any LOC given by \eqref{eq:equili} is an asymptotically stable equilibrium of the system.
\end{proof}
		
Theorem~\ref{th:stab} indicates that the closed-loop dynamic model of the multi-CSUR system in \eqref{eqn:z_k} asymptotically converges to a LOC with any initial conditions. Although multiple LOCs may exist and may not be globally stable, Theorem~\ref{th:stab} ensures the convergence of the system to a certain LOC, implying the achievement of objective \ref{it:pb:equi}) of problem~\ref{pb:unictr}.

Three control parameters are important to the proposed coverage controller in \eqref{eq:u_k}. For all $k\in \mathcal{N}$, $\gamma_k$ is the control gain that adjusts the amplitude of the control input, $\delta_k$ is the boundary layer scalar that smooths up the control inputs in zero vicinity, and ${Q_k}$ is a gain matrix that tunes the coverage cost function. Increasing $\gamma_k$, ${Q_k}$ and decreasing $\delta_k$ speeds up the convergence of the system state to a LOC. A simulation case study on how these control parameters affect the system performance will be presented in Sec.~\ref{sec:exp_param}.

\subsection{Distributed Control Implementation}\label{sec:dci}

\begin{figure}[htbp]
\centering
\noindent

\begin{tikzpicture}[scale=1,font=\small]

\def\nw{1cm}
\def\nh{1cm}

\definecolor{s_pink}{RGB}{255, 153, 153}
\definecolor{s_blue}{RGB}{153, 204, 255}
\definecolor{s_yellow}{RGB}{255, 230, 153}

\node[minimum height=\nh,minimum width=\nw, text width =0.8*\nw,align=center,draw,thick] (controller) at (0.2cm,2cm) {\eqref{eq:u_k}};

\node[minimum height=\nh,minimum width=1.4*\nw,text width =1.4*\nw,align=center,draw,thick] (agent) at (2.5cm,2cm) {Agent $k$\\\eqref{eq:virtdy}};

\node[minimum height=\nh,minimum width=\nw,text width =0.8*\nw,align=center,draw,thick] (v1) at (-2.7cm,2cm) {\eqref{eq:partialVz2}};

\node[minimum height=0.1cm,minimum width=2cm,inner sep=0pt,draw,fill=black] (bar) at (-2.7cm, 1.1cm) {};

\node[minimum height=2.4*\nh,minimum width=4.4*\nw,dashed,draw,very thick] (dashbox) at (-1cm,-0.7cm) {};

\node[minimum height=0.6\nh,minimum width=4*\nw,align=center,draw,thick] (net) at (0cm,-3.2cm) {Local communication among adjacent agents};

\node[minimum height=\nh,minimum width=\nw,text width =0.8*\nw,align=center,draw,thick] (c2) at (0.5cm,-1.2cm) {\eqref{eq:dp}};

\node[minimum height=\nh,minimum width=\nw,text width =0.8*\nw,align=center,draw,thick] (c1) at (-2.5cm,-1.2cm) {\eqref{eq:detC}};

\node[align=center, anchor=north east] () at (dashbox.north east) {$\forall\,i \in \mathscr{A}_k$};

\draw[->,>=stealth, very thick] (bar.north) -- (v1.south);

\draw[->,>=stealth,thick] (v1.east) -- node[pos=0.5,align=center, anchor=north]{$\displaystyle \nabla_k V(\mathcal{Z})$} (controller.west);

\draw[->,>=stealth,thick] (controller.east) -- node[pos=0.5,align=center, anchor=south]{$u_k(t)$} (agent.west);

\draw[->,>=stealth,thick] (agent.east) -- ([xshift=0.5cm] agent.east) -- ([xshift=0.5cm, yshift=-2cm] agent.east) -- node[pos=0.25,align=right, anchor=east, text width =1.5cm]{$z_k$, $\mathscr{A}_k$\\ $M(\mathcal{Z}_{\overline{\mathscr{A}_k}})$\\ $C(\mathcal{Z}_{\overline{\mathscr{A}_k}})$}([xshift=0.5cm, yshift=-5.2cm] agent.east) -- (net.east);

\draw[->,>=stealth,thick] (c1.north) -- node[pos=0.3,align=center, anchor=west]{$\displaystyle \nabla_k C(\mathcal{Z}_{\!\overline{\mathscr{A}_i}})$} ([xshift=0.2cm] bar.south);

\draw[->,>=stealth,thick] (c2.west) -- node[pos=0.5,align=center, anchor=center]{$D(\mathcal{Z}_{\overline{\mathscr{A}_i} }, z_k)$\\$P(\mathcal{Z}_{\overline{\mathscr{A}_i} }, z_k)$} (c1.east);

\draw[->,>=stealth,very thick] (net.west) -- ([xshift=-0.8cm, yshift=-4.25cm] bar.south) --node[pos=0.17,align=left, anchor=west]{$z_i, \mathscr{A}_i, M(\mathcal{Z}_{\overline{\mathscr{A}_i}}), C(\mathcal{Z}_{\overline{\mathscr{A}_i}}),~\forall\, i \in \overline{\mathscr{A}_k}$} ([xshift=-0.8cm, yshift=-0.3cm] bar.south) -- ([xshift=-0.2cm, yshift=-0.3cm] bar.south) -- ([xshift=-0.2cm] bar.south);

\draw[->,>=stealth, very thick] ([xshift=-0.8cm, yshift=-3.2cm] bar.south) -- ([yshift=-0.45cm] c2.south) -- (c2.south);

\draw[->,>=stealth, very thick] ([xshift=-0.8cm, yshift=-3.2cm] bar.south) -- ([yshift=-0.45cm] c1.south) -- (c1.south);

\draw[->,>=stealth, thick] (agent.south) -- ([yshift=-1cm] agent.center) -- node[pos=0.5,align=center, anchor=south]{$\theta_k$} ([yshift=-1cm] controller.center) -- (controller.south);

\end{tikzpicture}
\caption{The distributed control of each agent $k \in \mathcal{N}$.}
\label{fig:diag}
\end{figure}
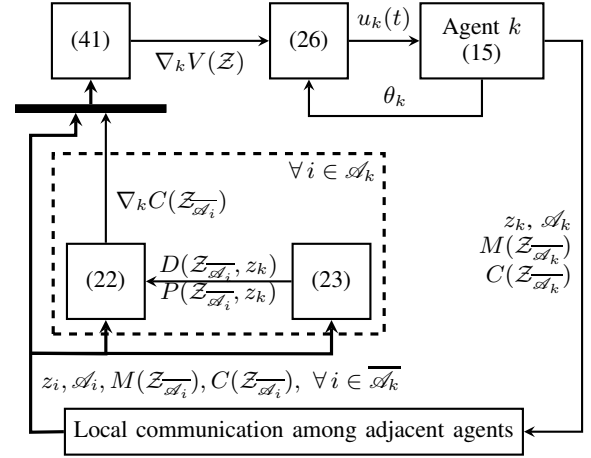

Now, we showcase that our coverage controller in \eqref{eq:u_k} can be implemented distributedly via a novel measurement-based method. A measurement-based method defines what information should be shared in the local communication among adjacent agents~\cite{ma2010consensus}.
Applying Proposition \ref{eq:distrib} to the cost gradient $\nabla_k V(\mathcal{Z})$ in \eqref{eq:partialVz}, we rewrite it as 
\begin{equation}\label{eq:partialVz2}
\begin{split}
\nabla_k V(\mathcal{Z}) =\, &\sum_{j=1}^M \left( \frac{{Q_k}(z_k - C(\mathcal{Z}_{\!\overline{\mathscr{A}_k}}))}{h_j(z_k)} + \frac{a_jW(\mathcal{Z}_{\!\overline{\mathscr{A}_k}})}{h^2_j(z_k)} \right. \\
& -\!\sum_{i \in \overline{\mathscr{A}_k}} \!\! \left. \nabla_k C(\mathcal{Z}_{\!\overline{\mathscr{A}_i}}) \frac{ {Q_k}(z_i-C(\mathcal{Z}_{\!\overline{\mathscr{A}_i}}))}{h_j(z_i)} 
\right),
\end{split}
\end{equation}
for $\mathcal{Z}_{\!\overline{\mathscr{A}_i}} \in \mathrm{int} \, \Omega^{|\!\overline{\mathscr{A}_i}|}$.
This implies that $\displaystyle \nabla_k V(\mathcal{Z})$ for $k \in \mathcal{N}$ only needs the information from agent $k$ and its adjacent agents $i \in \mathscr{A}_k$. Thus, it is possible to implement the controller in \eqref{eq:u_k} in a distributed manner by substituting $\displaystyle \nabla_k V(\mathcal{Z})$ with \eqref{eq:partialVz2}. Due to the nontrivial CSUR dynamic models, the shared measurement is more complicated than SIRs with only agent positions involved. Specifically, the calculation of the control input \eqref{eq:u_k} for each CSUR agent $k \in \mathcal{N}$ needs the following measurements of its adjacent agents $i \in \mathscr{A}_k$: 
1) position $z_i$, 
2) Voronoi mass $M(\mathcal{Z}_{\overline{\mathscr{A}_i}})$ and centroid $C(\mathcal{Z}_{\overline{\mathscr{A}_i}})$,
and 3) adjacency relation $\mathscr{A}_i$ used to determine $\partial \Omega_k^i$ and calculate $D(\mathcal{Z}_{\overline{\mathscr{A}_i} }, z_k)$ and $P(\mathcal{Z}_{\overline{\mathscr{A}_i} }, z_k)$.
Our measurement-based method for distributed coverage control is illustrated in Fig.~\ref{fig:diag}.



The main computational load of the controller in \eqref{eq:u_k} comes from the calculation of the gradient $\displaystyle \nabla V(\mathcal{Z})$ which requires the centroid gradients $\displaystyle \nabla_k C(\mathcal{Z}_{\!\overline{\mathscr{A}_i}})$ for $i,k\!\in\!\mathcal{N}$. The distributed form in \eqref{eq:partialVz2} only needs the computation of $\displaystyle \nabla_k C(\mathcal{Z}_{\!\overline{\mathscr{A}_i}})$ among adjacent agents $i \in \overline{\mathscr{A}}_k$, instead of all agent pairs $i,k\!\in\!\mathcal{N}$. This leads to a greatly reduced computational load compared to the centralized implementation using \eqref{eq:partialVz}, considering that the value of $|\overline{\mathscr{A}_{k}}|$ for each $k\in \mathcal{N}$ is typically much smaller than the total agent number $N$. This allows the distributed coverage controller to be implemented in real-time even for large-scale systems.


During the convergence to a LOC, the adjacency relation $\mathscr{A}_i$ for any agent $i\!\in\!\mathcal{N}$ may be time-variant~\cite{cortes2005coordination}. 
Previous work on the distributed characterization of Voronoi partitions~\cite{hadjicostis2003distributed, elwin2017distributed} can be used to update time-variant adjacency relations. This facilitates a fully distributed measurement-based method for distributed coverage control.

\section{Simulation Studies} \label{sec:sim}

We validate the performance of the proposed coverage controller in simulation studies.
We first validate the efficacy of the controller for six CSUR agents with various initial conditions and control parameters. Then, we verify the scalability of the controller for a larger system with 100 agents. Finally, a comparison study with the conventional coverage controller is performed to address the advantage of the proposed method in avoiding infeasibility. All studies are simulated in MATLAB R2021a at a discrete sampling time $0.05\,$s.

\subsection{Method Test with Different Initial Conditions}\label{sec:init}

\begin{figure*}[htbp]
     \centering
     \begin{subfigure}[b]{0.32\textwidth}
         \centering
         \includegraphics[width=\textwidth]{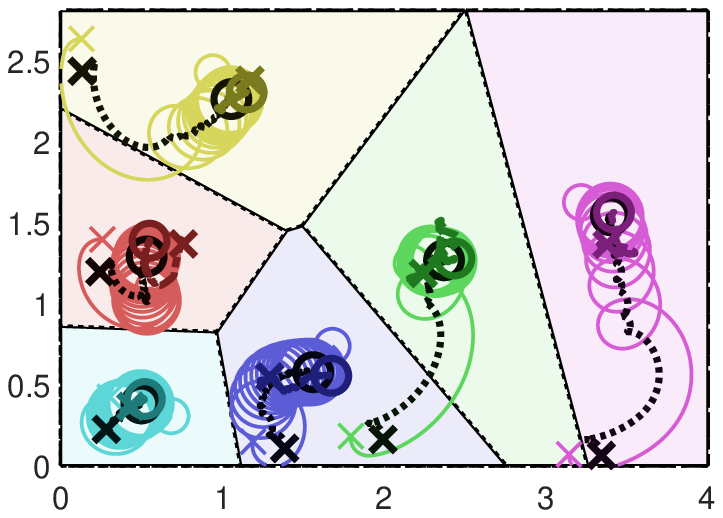}
         \caption{The agent trajectories of Case \#\,1}
         \label{fig:init_trajectories_1}
     \end{subfigure}
     \hfill
     \begin{subfigure}[b]{0.32\textwidth}
         \centering
         \includegraphics[width=\textwidth]{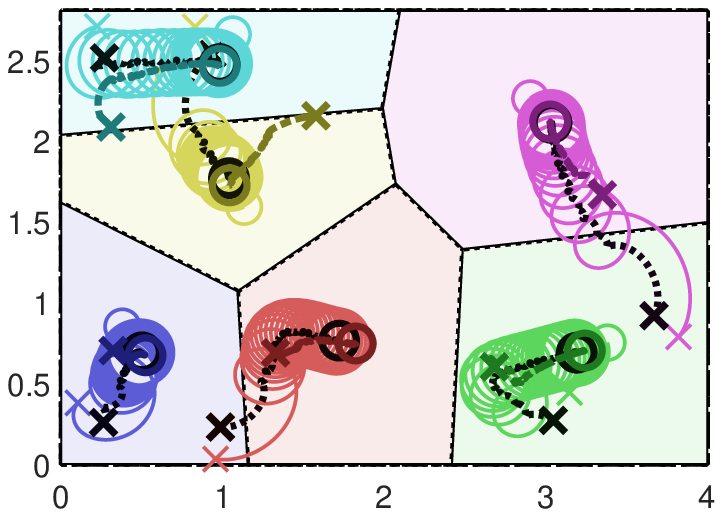}
         \caption{The agent trajectories of Case \#\,2}
         \label{fig:init_trajectories_2}
     \end{subfigure}
     \hfill
     \begin{subfigure}[b]{0.32\textwidth}
         \centering
         \includegraphics[width=\textwidth]{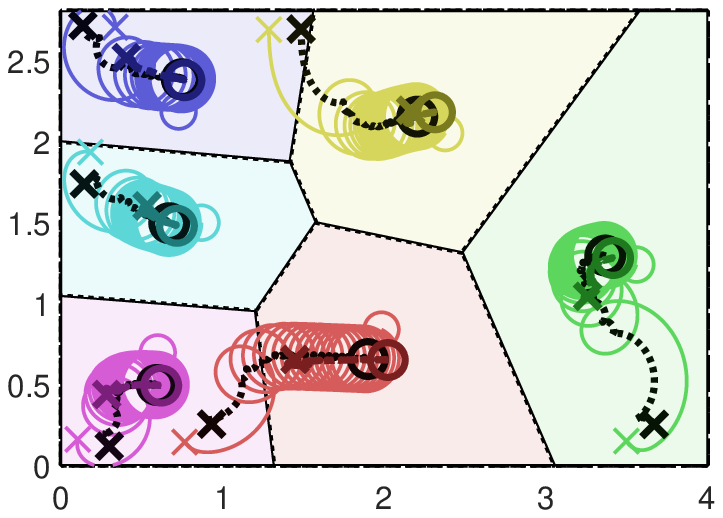}
         \caption{The agent trajectories of Case \#\,3}
         \label{fig:init_trajectories_3}
     \end{subfigure}
     \hfill
     \begin{subfigure}[b]{0.32\textwidth}
         \centering
         \includegraphics[width=\textwidth]{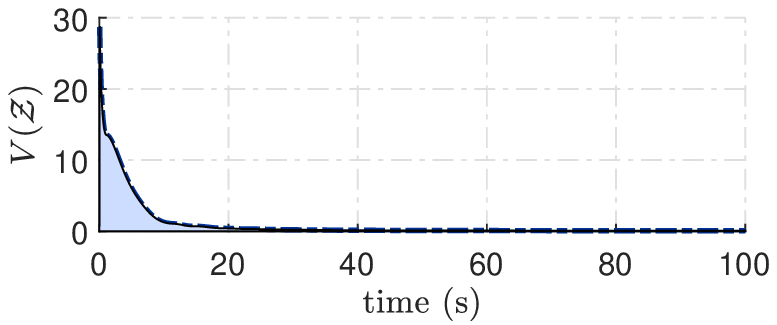}
         \caption{The coverage cost of Case \#\,1}
         \label{fig:init_cost_1}
     \end{subfigure}
     \hfill
     \begin{subfigure}[b]{0.32\textwidth}
         \centering
         \includegraphics[width=\textwidth]{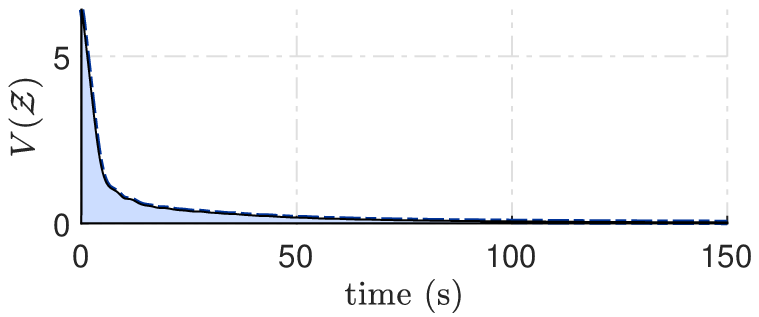}
         \caption{The coverage cost of Case \#\,2}
         \label{fig:init_cost_2}
     \end{subfigure}
     \hfill
     \begin{subfigure}[b]{0.32\textwidth}
         \centering
         \includegraphics[width=\textwidth]{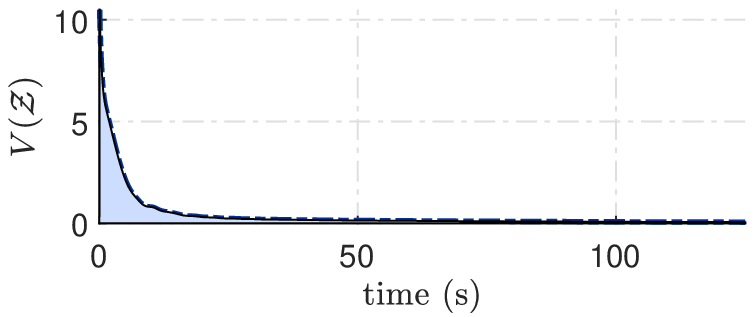}
         \caption{The coverage cost of Case \#\,3}
         \label{fig:init_cost_3}
     \end{subfigure}
     \hfill
     \begin{subfigure}[b]{0.32\textwidth}
         \centering
         \includegraphics[width=\textwidth]{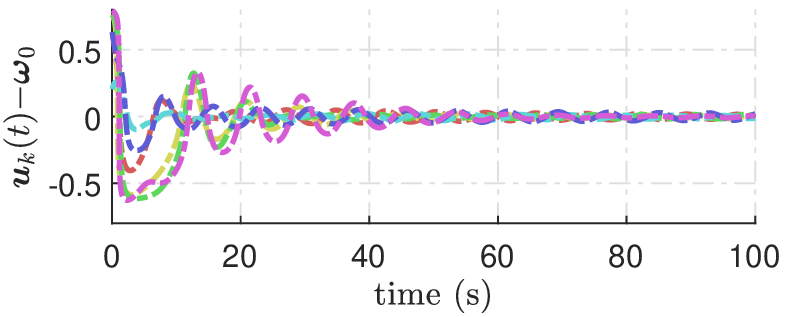}
         \caption{The control inputs of Case \#\,1}
         \label{fig:init_input_1}
     \end{subfigure}
     \hfill
     \begin{subfigure}[b]{0.32\textwidth}
         \centering
         \includegraphics[width=\textwidth]{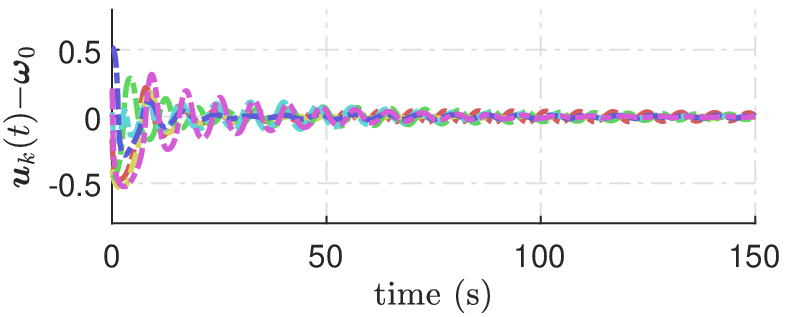}
         \caption{The control inputs of Case \#\,2}
         \label{fig:init_input_2}
     \end{subfigure}
     \hfill
     \begin{subfigure}[b]{0.32\textwidth}
         \centering
         \includegraphics[width=\textwidth]{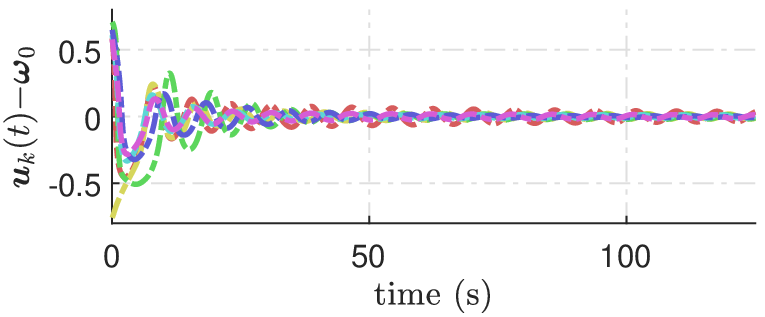}
         \caption{The control inputs of Case \#\,3}
         \label{fig:init_input_3}
     \end{subfigure}
        \caption{Simulation results in different initial conditions: (a)-(c) are CSUR positions $\zeta_k(t)$ (thin solid lines), virtual centers $z_k(t)$ (thick dotted lines), and Voronoi centroids $C(\mathcal{Z}_{\overline{\mathscr{A}_k}})$ (thick dashed lines), where `x' and `o' are the starting and ending points of the trajectories; (d)-(f) are the coverage costs $V(\mathcal{Z})$, and (g)-(i) are the control inputs $u_k(t) - \omega_k$..}
        \label{fig:init}
\end{figure*}

This study tests the performance of the proposed method for six CSUR agents with different initial conditions. The target region $\Omega$ is a $4\,$m $\times$ $2.8\,$m rectangular region. The boundary functions $h_j(\omega)$, $j=1,2,3,4$, $\omega \in \Omega$, are parameterized by $a_1 \!=\! [\,-1~0\,]$, $b_1 \!=\! 0$, $a_2 \!=\! [\,1~0\,]$, $b_2 \!=\! 4$, $a_3 \!=\! [\,0~1\,]$, $b_3 = 2.8$, $a_4 \!=\! [\,0~-1\,]$, $b_4 \!=\! 0$. The linear speed and the nominal angular velocity of the CSURs are set as identical values $v_k \!=\! 0.16\,$m/s and $\omega_k \!=\! 0.8\,$rad/s, for all $k\!=\!1,2,\cdots,6$, for simplicity. Three different initial configurations are randomly generated and assigned to the CSURs, as shown in Tab.~\ref{tab:param}, where $[\,\zeta_x~\zeta_y\,]^{\!\top\!}$ and $\theta$ are the planar coordinate and the orientation of a CSUR. For all cases and all agents $k\in \mathcal{N}$, the control parameters are selected as $\gamma_k \!=\! 1$, $Q_k \!=\! I$, and $\delta_k\!=\!2$.

\linespread{1.2}
\begin{table}[htbp]
\centering
\caption{The Initial Configurations of Cases \#$\,$1, \#$\,$2, and \#$\,$3}
\begin{tabular}{c|c|cccccc}
	\hline
	\multicolumn{2}{c|}{\#$\,$Agent} & 1 & 2 & 3 & 4 & 5 & 6 \\
	\hline
	\multirow{3}{*}{\#$\,$1} &$\zeta_x$ & 0.2546 & 0.1247 & 1.793 & 0.3006 & 1.187 & 3.144 \\
& $\zeta_y$ & 1.392 & 2.629 & 0.1781 & 0.4191 & 0.1445 & 0.0658 \\
& $\theta$ & 3.060 & 3.160 &	4.610 & 3.030 & 4.500 &	4.680 \\
	\hline
	\multirow{3}{*}{\#$\,$2} &$\zeta_x$ & 0.9549 & 0.8286 & 3.148 & 0.2219 & 0.1023 & 3.823 \\
	& $\zeta_y$ & 0.0310 & 2.702 & 0.4426 & 2.705 & 0.3783 & 0.7863 \\
	&$\theta$ & 6.130 & 3.690 & 2.610 & 3.370 & 4.060 & 0.8600 \\
	\hline
	\multirow{3}{*}{\#$\,$3} &$\zeta_x$ & 0.8690 & 1.3810 & 3.610 & 0.7773 & 0.3674 & 0.4060 \\
	& $\zeta_y$ & 0.1436 & 2.6980 & 0.2723 & 2.726 & 2.610 & 0.2589 \\
	& $\theta$ & 4.760 & 4.560 & 4.390 & 4.650 & 1.430 & 1.340 \\
	\hline
\end{tabular}
\label{tab:param}
\end{table}
\linespread{1}

The results are illustrated in Fig.~\ref{fig:init}.
The trajectories of the robot positions, virtual centers, and Voronoi centroids are presented in Fig.~\ref{fig:init_trajectories_1}, Fig.~\ref{fig:init_trajectories_2}, and Fig.~\ref{fig:init_trajectories_3}, respectively. The trajectories of all agents (the virtual centers of the CSURs) are confined within the region for all time, which indicates the achievement of objective \ref{it:pb:state}) of Problem \ref{pb:unictr}. All virtual centers and their corresponding Voronoi centroids, both marked as `o' but with different colors, coincide with each other ultimately, which verifies the ultimate achievement of optimal coverage. The coinciding points indicate the corresponding LOC. The CSURs ultimately orbit around these points at a radius $50\,$m which allows a low likelihood of collisions. The achievement of optimal coverage is also reflected in Fig.~\ref{fig:init_cost_1}, Fig.~\ref{fig:init_cost_2}, and Fig.~\ref{fig:init_cost_3}, where the coverage function decays to zero within $100\,$s for all initial conditions. Besides, the control inputs of all robots shown in Fig.~\ref{fig:init_input_1}, Fig.~\ref{fig:init_input_2}, and Fig.~\ref{fig:init_input_3} are all strictly confined by $|u_k(t)-\omega_k| < \gamma_k \omega_k = 0.8$ for all agents $k=1,2,\cdots,6$ and all time, which achieves objective \ref{it:pb:input}) of Problem \ref{pb:unictr}. Note that different initial conditions ultimately lead to different LOCs. They may also affect the convergence speed of the coverage cost. Therefore, we can conclude that the proposed coverage controller \eqref{eq:u_k} achieves all three objectives of Problem~\ref{pb:unictr} with different initial conditions. 

\begin{figure*}[htbp]
     \centering
     \begin{subfigure}[b]{0.32\textwidth}
         \centering
         \includegraphics[width=\textwidth]{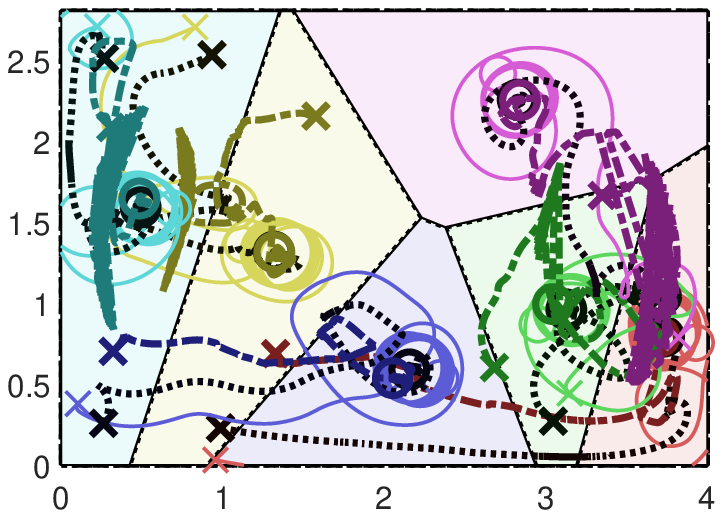}
         \caption{$\gamma_k = 10$, $Q_k=I$, $\delta_k=2$}
         \label{fig:param_trajectories_1}
     \end{subfigure}
     \hfill
     \begin{subfigure}[b]{0.32\textwidth}
         \centering
         \includegraphics[width=\textwidth]{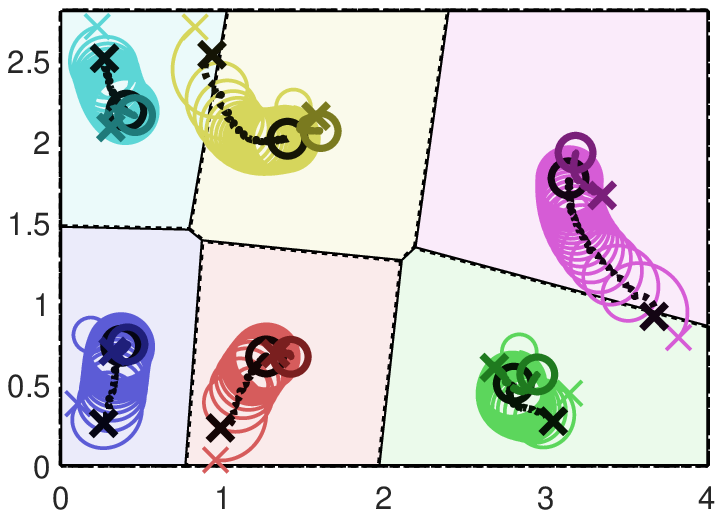}
         \caption{$\gamma_k = 1$, $Q_k=I$, $\delta_k=10$.}
         \label{fig:param_trajectories_2}
     \end{subfigure}
     \hfill
     \begin{subfigure}[b]{0.32\textwidth}
         \centering
         \includegraphics[width=\textwidth]{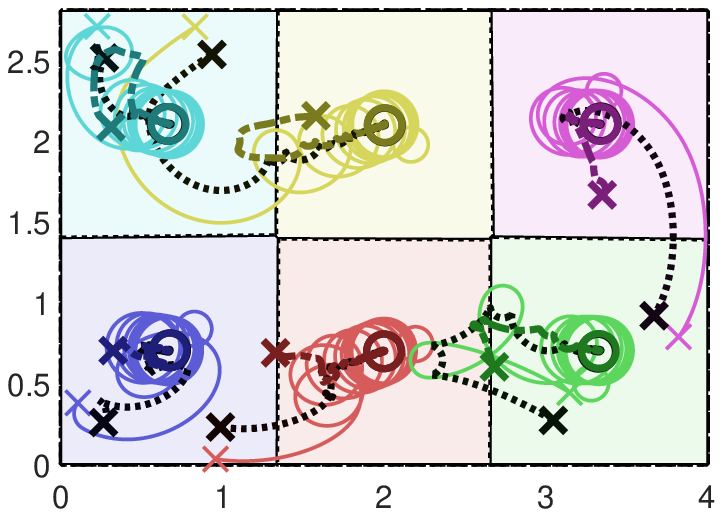}
         \caption{$\gamma_k = 1$, $Q_k=10I$, $\delta_k=2$}
         \label{fig:param_trajectories_3}
     \end{subfigure}
    \hfill
     \begin{subfigure}[b]{0.32\textwidth}
         \centering
         \includegraphics[width=\textwidth]{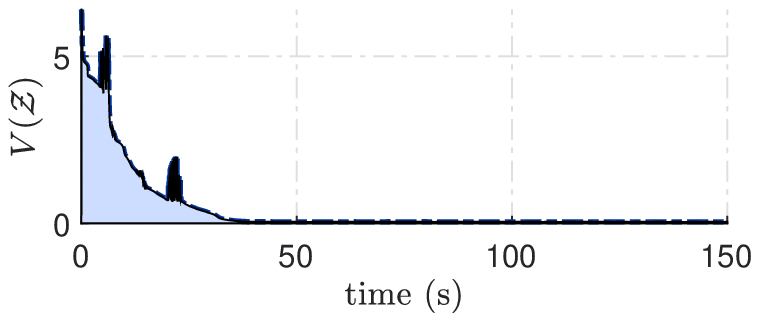}
         \caption{$\gamma_k = 10$, $Q_k=I$, $\delta_k=2$}
         \label{fig:param_cost_1}
     \end{subfigure}
     \hfill
     \begin{subfigure}[b]{0.32\textwidth}
         \centering
         \includegraphics[width=\textwidth]{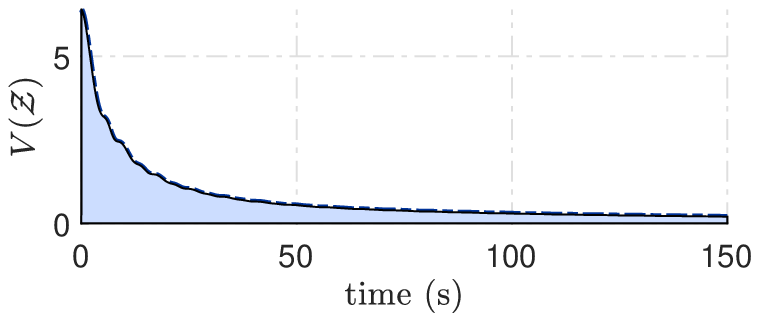}
         \caption{$\gamma_k = 1$, $Q_k=I$, $\delta_k=10$}
         \label{fig:param_cost_2}
     \end{subfigure}
     \hfill
     \begin{subfigure}[b]{0.32\textwidth}
         \centering
         \includegraphics[width=\textwidth]{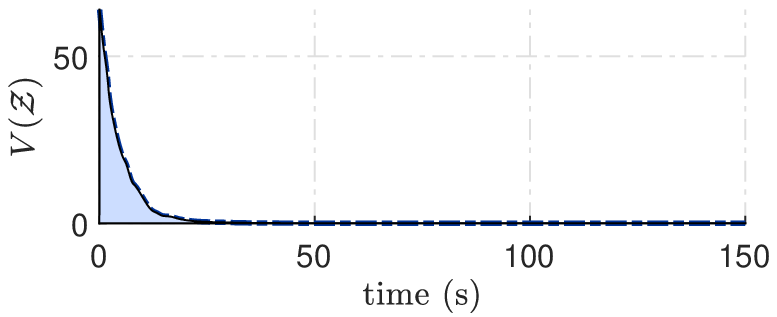}
         \caption{$\gamma_k = 1$, $Q_k=10I$, $\delta_k=2$}
         \label{fig:param_cost_3}
     \end{subfigure}
    \hfill
     \begin{subfigure}[b]{0.32\textwidth}
         \centering
         \includegraphics[width=\textwidth]{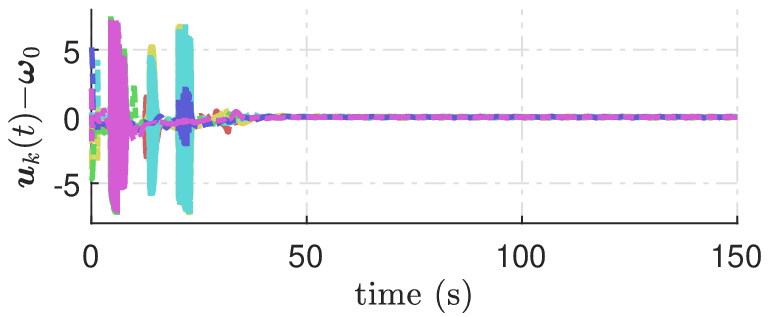}
         \caption{$\gamma_k = 10$, $Q_k=I$, $\delta_k=2$}
         \label{fig:param_input_1}
     \end{subfigure}
     \hfill
     \begin{subfigure}[b]{0.32\textwidth}
         \centering
         \includegraphics[width=\textwidth]{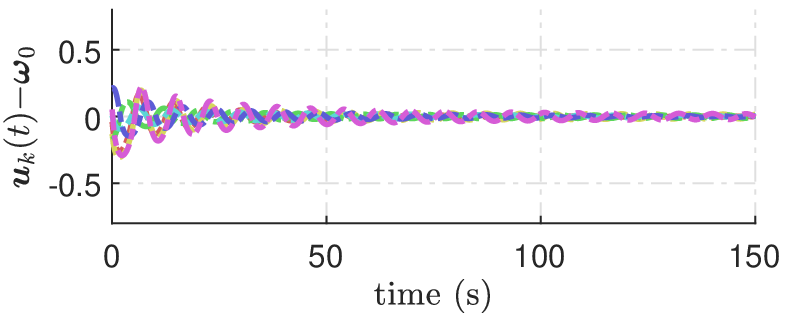}
         \caption{$\gamma_k = 1$, $Q_k=I$, $\delta_k=10$}
         \label{fig:param_input_2}
     \end{subfigure}
     \hfill
     \begin{subfigure}[b]{0.32\textwidth}
         \centering
         \includegraphics[width=\textwidth]{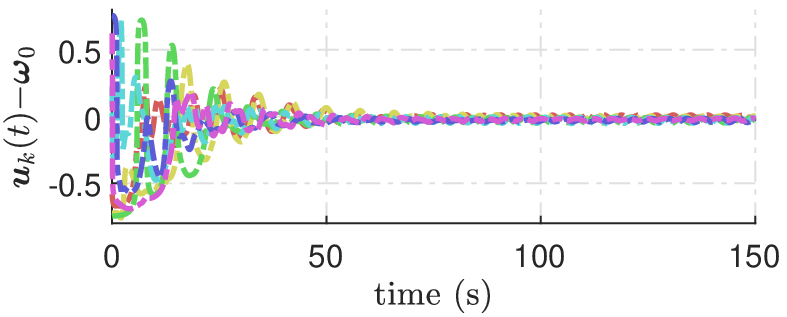}
         \caption{$\gamma_k = 1$, $Q_k=10I$, $\delta_k=2$}
         \label{fig:param_input_3}
     \end{subfigure}
        \caption{Simulation results with different control parameters: (a)-(c) are CSUR positions $\zeta_k(t)$ (thin solid lines), virtual centers $z_k(t)$ (thick dotted lines), and Voronoi centroids $C(\mathcal{Z}_{\overline{\mathscr{A}_k}})$ (thick dashed lines), where `x' and `o' are the starting and ending points of the trajectories, (d)-(f) are the coverage costs $V(\mathcal{Z})$, and (g)-(i) are the control inputs $u_k(t) - \omega_k$.}
        \label{fig:param}
\end{figure*}

\subsection{The Influence of the Control Parameters}\label{sec:exp_param}

This study evaluates the influence of the control parameters, namely the input gain $\gamma_k$, the coverage gain $Q_k$, and the boundary layer scalar $\delta_k$, on the performance of the proposed coverage controller, $k=1,2,\cdots,6$. The size of the target region and the robot parameters $v_k$, $\omega_k$ are the same as those in Sec.~\ref{sec:init}. The initial conditions of the agents are determined as Case \#\,2 in Tab.~\ref{tab:param}. The simulation results with different control parameters are illustrated in Fig.~\ref{fig:param}. We also compare the simulation results in Fig.~\ref{fig:param} with Case \#\,2 of Fig.~\ref{fig:init} since they have the same initial conditions. Similar to Sec.~\ref{sec:init}, Fig.~\ref{fig:param} indicates that optimal coverage is achieved for all cases with the trajectories of the virtual centers confined within the target region. All control inputs are restricted by $|u_k(t) -\omega_k| \!<\! \gamma_k \omega_k$, although the bounds are different. Thus, we can conclude that the proposed coverage controller \eqref{eq:u_k} well solves Problem~\ref{pb:unictr} with different control parameters.

Comparing Fig.~\ref{fig:param} with Case \#\,2 in Fig.~\ref{fig:init}, we notice that these parameters affect the control performance differently. Firstly, a large $\gamma_k$ increases the convergence rate of the coverage cost but also causes chattering to the control inputs. This is because the system tends to become unstable as the control gain becomes over-large due to the discrete sampling. Secondly, an over-large $\delta_k$ may decelerate the convergence to a LOC. Thirdly, a large $Q_k$ can effectively increase the convergence rate of the coverage cost without causing chatting to the control inputs. Thus, we suggest only using $\gamma_k$ to restrict the control inputs while increasing the value of the coverage gain $Q_k$ to improve the convergence rate. The scalar $\delta_k$ should be small to maintain a decent convergence rate while ensuring the smoothness of the control inputs.

\subsection{Optimal Coverage of A Larger-Scale System}\label{sec:large}
	
\begin{figure}[ht]
     \centering
     \begin{subfigure}[b]{0.45\textwidth}
         \centering
         \includegraphics[width=\textwidth]{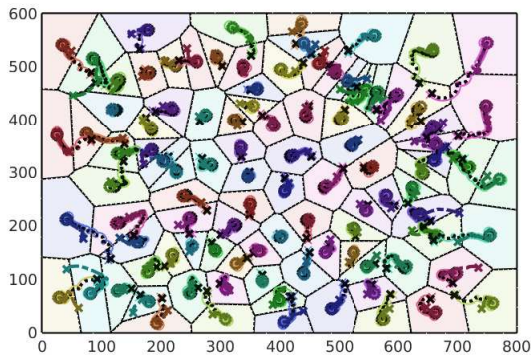}
         \caption{The CSUR positions $\zeta_k(t)$ (thin solid lines), virtual centers $z_k(t)$ (thick dotted lines), and Voronoi centroids $C(\mathcal{Z}_{\overline{\mathscr{A}_k}})$ (thick dashed lines) of the CSURs, $k\!=\!1,2,\cdots,100$, where `x' and `o' are the starting and ending points of the trajectories.}
         \label{fig:100agent_trajectories}
     \end{subfigure}
     \hfill
     \begin{subfigure}[b]{0.4\textwidth}
         \centering
         \includegraphics[width=\textwidth]{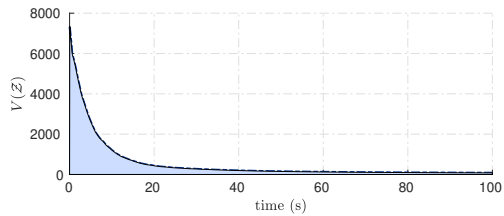}
         \caption{The value of the coverage cost as time changes}
         \label{fig:100agent_cost}
     \end{subfigure}
     \hfill
     \begin{subfigure}[b]{0.4\textwidth}
         \centering
         \includegraphics[width=\textwidth]{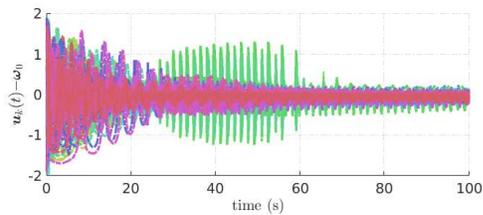}
         \caption{The control inputs as time changes}
         \label{fig:100agent_input}
     \end{subfigure}
        \caption{Optimal coverage of a 100-agent robot team.}
        \label{fig:24agent}
\end{figure}

This study tests the proposed coverage controller on a larger-scale multi-agent system that contains 100 CSURs. The coverage is performed on a $800\,$m$\times 600\,$m rectangular region with the same boundary coefficients as Sec.~\ref{sec:init}, except that $b_2 = 800$ and $b_3 = 600$. The linear speed and the nominal angular velocity of the CSURs are $v_k = 10\,$m/s and $\omega_k = 2\,$rad/s, $k=1,2,\cdots,100$, which correspond to a small orbit radius $5\,$m such that the CSURs are not likely to collide with each other. The control parameters are selected as $\gamma_k = 1$, $Q_k = 10\,I$, and $\delta_k=2$ for all agents. The initial positions of the robots are randomly sampled from the target region and are not listed here. The simulation results are illustrated in Fig.~\ref{fig:24agent} from which we can draw similar conclusions to the simulation study in Sec.~\ref{sec:init}. Specifically, from Fig.~\ref{fig:100agent_trajectories}, we can see that the virtual centers of all CSURs ultimately coincide with the Voronoi centroids. Also, Fig.~\ref{fig:100agent_cost} shows that the coverage cost decays to zero. Both subfigures indicate the success of the optimal coverage after around $60\,$s. This means that objective 3) of Problem \ref{pb:unictr} is achieved for this 100-agent system. The state-dependent and input-dependent constraints are also strictly satisfied. Fig.~\ref{fig:100agent_trajectories} indicates that all virtual centers are strictly confined within the covered region, and the control inputs are limited by $|u_k(t)-\omega_k| < \gamma_k \omega_k=2$ according to Fig.~\ref{fig:100agent_input}. Thus, objectives 1) and 2) of Problem \ref{pb:unictr} are also achieved. We can conclude that the proposed coverage controller is also effective for a large-scale multi-CSUR system that contains as many as 100 agents.

\subsection{A Comparison Study With the Conventional Method}\label{sec:comp}
	
\begin{figure}[ht]
     \centering
     \begin{subfigure}[b]{0.24\textwidth}
         \centering
         \includegraphics[width=\textwidth]{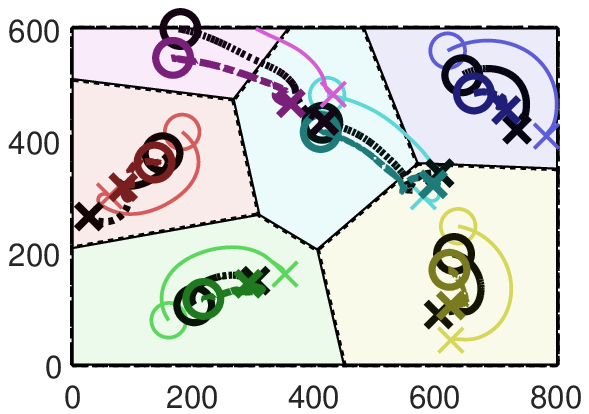}
         \caption{Conventional controller}
         \label{fig:compare_conv}
     \end{subfigure}
     \hfill
     \begin{subfigure}[b]{0.24\textwidth}
         \centering
         \includegraphics[width=\textwidth]{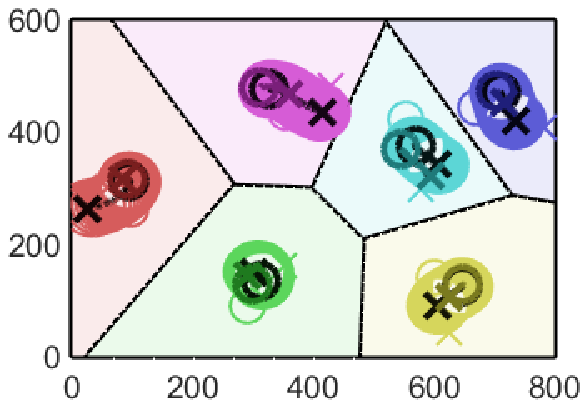}
         \caption{Proposed controller}
         \label{fig:compare_proposed}
     \end{subfigure}
        \caption{The CSUR positions $\zeta_k(t)$ (thin solid lines), the virtual centers $z_k(t)$ (thick dotted lines), and the Voronoi centroids $C(\mathcal{Z}_{\mathscr{A}_k})$ (thick dashed lines) of the multi-CSUR system with the \textit{conventional} and the \textit{proposed} coverage control methods.}
        \label{fig:compare}
\end{figure}

As mentioned in Sec.~\ref{sec:main}, the main advantage of our proposed coverage controller~\eqref{eq:u_k} over the conventional gradient-based controller~\eqref{eq:grad} is the additional state-dependent constraints~\eqref{eq:inv} that are critical to solving the feasibility issue for a multi-CSUR system. This subsection conducts a comparison study between these two methods to address the advantage of the proposed coverage controller. The detailed formulation of the conventional coverage controller is provided in~\cite{liu2017coverage}, of which the closed-loop dynamic model is $\displaystyle \dot{z}_k(t) = - \gamma \nabla_k H(\mathcal{Z})$, where $\displaystyle \nabla_k H(\mathcal{Z})$ is calculated using \eqref{eq:gradient}. This study is conducted in a $800\,$m$\times 600\,$m rectangular region with six CSURs. To showcase the advantage of our proposed controller compared to the conventional method in a fair comparison condition, for both controllers, we set the same velocity constants $v_k = 40\,$m/s and $\omega_k = 0.8\,$rad/s, the same initial positions as shown in Tab.~\ref{tab:param_comp}, and the same control gain $\gamma_k=0.1$ for all $k=1,2,\cdots,6$. The only difference is that the conventional controller uses a conventional coverage cost function $H(\mathcal{Z})$ in~\eqref{eqn:Hprimitive} with $\Phi(\omega)=1$, $\omega \in \Omega$, but the proposed controller uses the novel coverage cost function in~\eqref{eqn:V} with $Q_k = I$ and $\delta_k=2$ for all agents $k=1,2,\cdots,6$. The trajectories of the CSUR positions, virtual centers, and Voronoi centroids are illustrated in Fig.~\ref{fig:compare}. Fig.~\ref{fig:compare_conv} clearly shows that one virtual center tends to cross the region boundary and move outside the target region while the optimal coverage has not been reached. The situation after this is not drawn since the Voronoi partition is no longer feasible. However, the proposed controller ensures that all virtual centers are confined within the target region and ultimately coincide with the Voronoi centroids, as shown in Fig.~\ref{fig:compare_proposed}. This implies that the proposed controller ensures the feasibility of the optimal coverage problem but the conventional method does not under the same conditions. Since the feasibility issue is the major challenge this work focuses on, other metrics are not considered in this comparison study. 

\linespread{1.2}
\begin{table}[htbp]
\centering
\caption{The Initial Condition of the Comparison Study}
\begin{tabular}{c|cccccc}
	\hline
	\#$\,$Agent & 1 & 2 & 3 & 4 & 5 & 6 \\
	\hline
	$\zeta_x$ & 60.68 & 624.4 & 350.6 & 579.2 & 782.5 & 430.3 \\
	 $\zeta_y$ & 301.0 & 43.43 & 161.5 & 299.7 & 408.0 & 482.4 \\
	 $\theta$ & 2.394 & 0.414 & 1.810 & 5.715 & 1.341 & 2.841 \\
	\hline
\end{tabular}
\label{tab:param_comp}
\end{table}
\linespread{1}

\section{Experiment Validation} \label{sec:exp}

\begin{figure}
 \centering
 \hspace{-0.85cm}
 \begin{subfigure}[b]{0.36\textwidth}
     \centering
     \includegraphics[height=3cm]{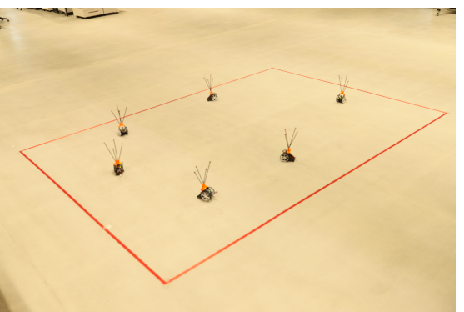}
 \caption{The experimental ground region.}
 \label{fig:ground}
 \end{subfigure}
 \hspace{-0.5cm}
 \begin{subfigure}[b]{0.12\textwidth}
     \centering
     \includegraphics[height=3cm]{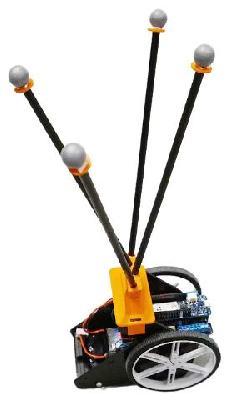}
 \caption{Robot}
 \label{fig:robot}
 \end{subfigure}
 \caption{The experimental setup and the mobile robot.}
 \label{fig:exp}
\end{figure}

In this section, we conduct an experimental study on real robot platforms to verify the applicability of the proposed method. The target region is a $4\,$m $\times 2.8\,$m indoor area, as shown in Fig.~\ref{fig:ground}. We use six two-wheel unicycle mobile robots provided by the Arduino Engineering Kit\textregistered, as shown in Fig.~\ref{fig:robot}, to serve as the CSURs. Each robot is attached with four infra-tracking markers allowing its motion to be tracked by a Qualisys\textregistered~motion tracking system at a frequency of 300 Hz. A Lenovo Thinkpad laptop with an Intel core I5-6200U CPU and 8GB RAM, running with the Ubuntu 16.04 operating system, receives the measurements from the tracking system and sends control commands to the robots. Each robot is controlled by a robotic operating system (ROS) node on the laptop at an update frequency of 100 Hz. The desired linear speed $v_k \!=\! 0.16\,$m/s and angular velocity $\omega_k \!=\!0.8\,$rad/s are converted to the control commands for the wheel motors of all robots $k=1,2,\cdots,6$. A PD controller is used to ensure the robots maintain these speeds. The motion tracking system, the laptop, and the mobile robots are connected using a wireless network. The adjacency relation among the robots is computed using the distributed algorithm introduced in~\cite{hadjicostis2003distributed, elwin2017distributed}. 

Note that the ROS network used to coordinate the control and measurement of the robot is not subject to hard real-time and does not ensure constant discrete sampling. Also, a communication delay exists in the network due to its limited bandwidth. Moreover, the robots' linear speed and angular velocity are not ideally constant due to the friction forces and their motor features. All these factors lead to uncertainties in the experiment. Therefore, the main purpose of this experiment is to investigate the difference between the experiment and simulation results under the same conditions and evaluate how the uncertainties affect the performance of the proposed coverage controller. For a fair comparison, the initial conditions and the control parameters in this experiment are the same as the simulation study in Sec.~\ref{sec:init}.

\begin{figure*}[htbp]
 \centering
 \begin{subfigure}[b]{0.32\textwidth}
     \centering
     \includegraphics[width=\textwidth]{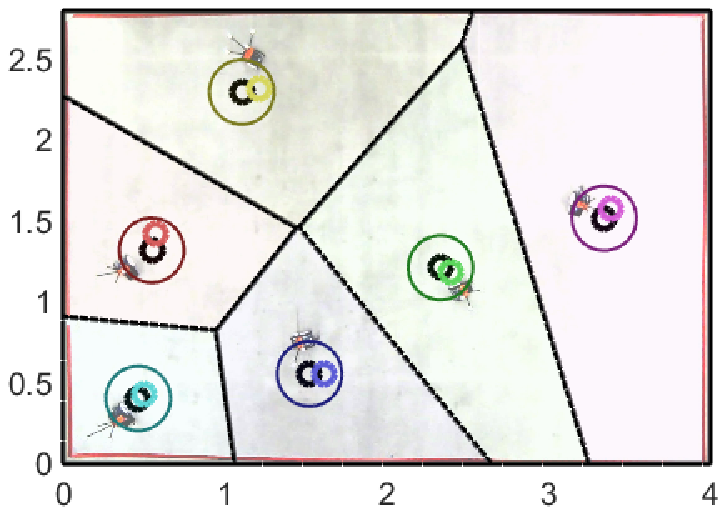}
     \caption{The agent trajectories of Case \#\,1}
     \label{fig:exp_image_1}
 \end{subfigure}
 \hfill
 \begin{subfigure}[b]{0.32\textwidth}
     \centering
     \includegraphics[width=\textwidth]{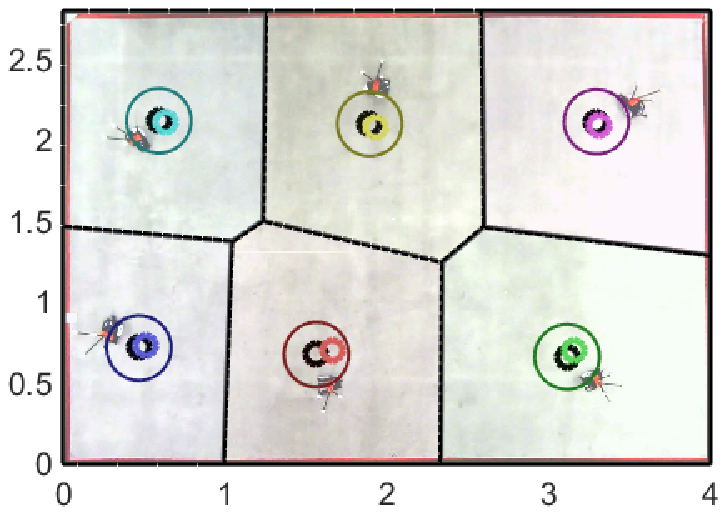}
     \caption{The agent trajectories of Case \#\,2}
     \label{fig:exp_image_2}
 \end{subfigure}
 \hfill
 \begin{subfigure}[b]{0.32\textwidth}
     \centering
     \includegraphics[width=\textwidth]{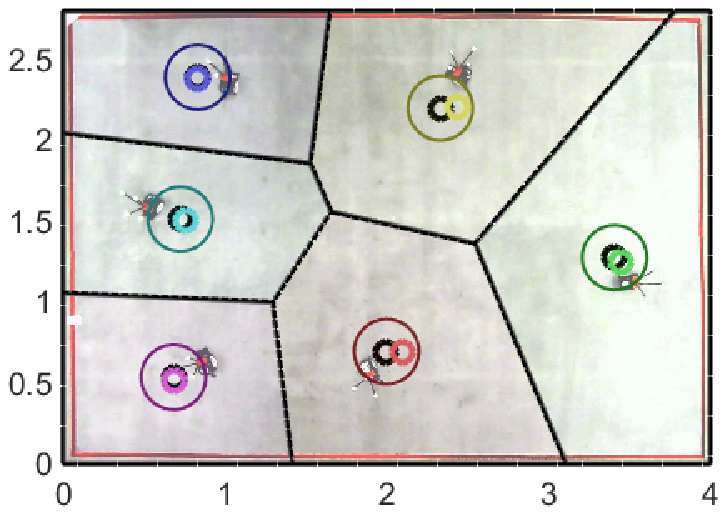}
     \caption{The agent trajectories of  Case \#\,3}
     \label{fig:exp_image_3}
 \end{subfigure}
 \hfill
 \begin{subfigure}[b]{0.32\textwidth}
     \centering
     \includegraphics[width=\textwidth]{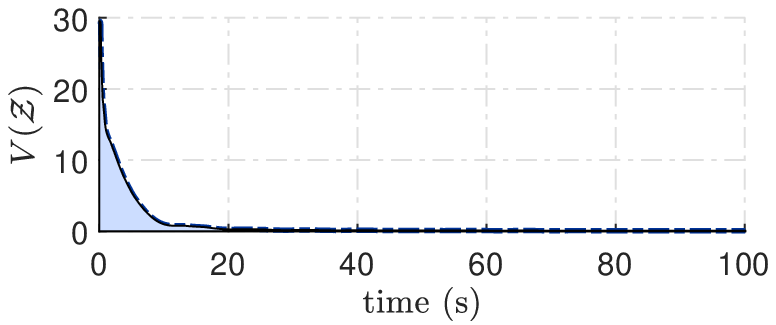}
     \caption{The coverage cost of Case \#\,1}
     \label{fig:exp_cost_1}
 \end{subfigure}
 \hfill
 \begin{subfigure}[b]{0.32\textwidth}
     \centering
     \includegraphics[width=\textwidth]{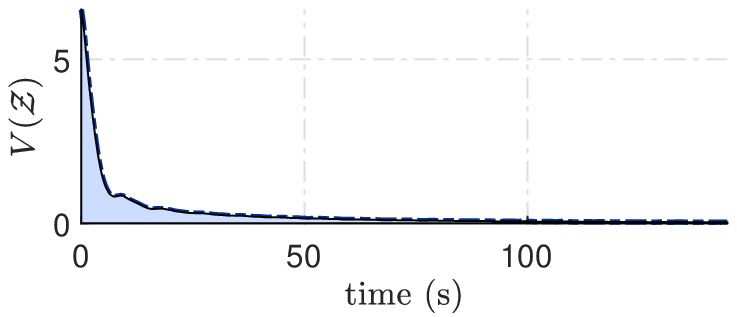}
     \caption{The coverage cost of Case \#\,2}
     \label{fig:exp_cost_2}
 \end{subfigure}
 \hfill
 \begin{subfigure}[b]{0.32\textwidth}
     \centering
     \includegraphics[width=\textwidth]{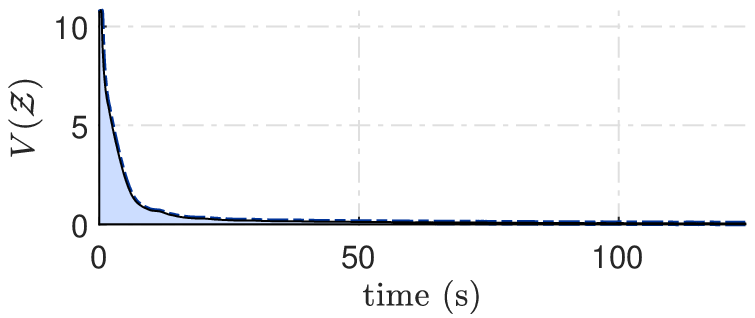}
     \caption{The coverage cost of Case \#\,3}
     \label{fig:exp_cost_3}
 \end{subfigure}
 \hfill
 \begin{subfigure}[b]{0.32\textwidth}
     \centering
     \includegraphics[width=\textwidth]{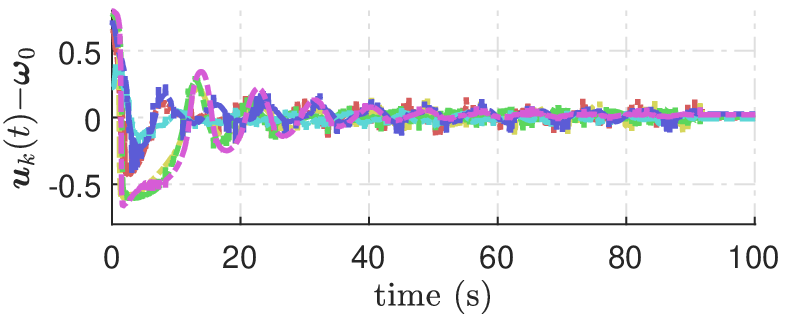}
     \caption{The cost inputs of Case \#\,1}
     \label{fig:exp_input_1}
 \end{subfigure}
 \hfill
 \begin{subfigure}[b]{0.32\textwidth}
     \centering
     \includegraphics[width=\textwidth]{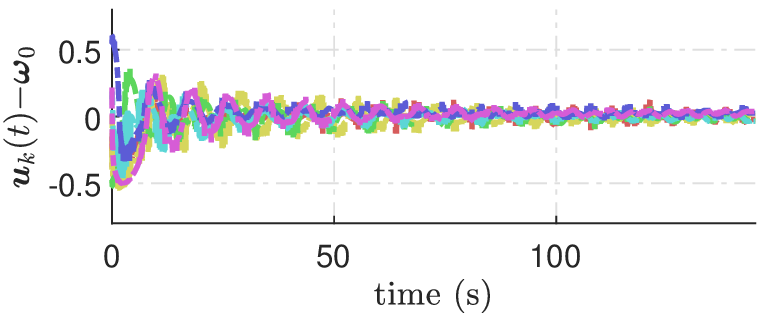}
     \caption{The cost inputs of Case \#\,2}
     \label{fig:exp_input_2}
 \end{subfigure}
 \hfill
 \begin{subfigure}[b]{0.32\textwidth}
     \centering
     \includegraphics[width=\textwidth]{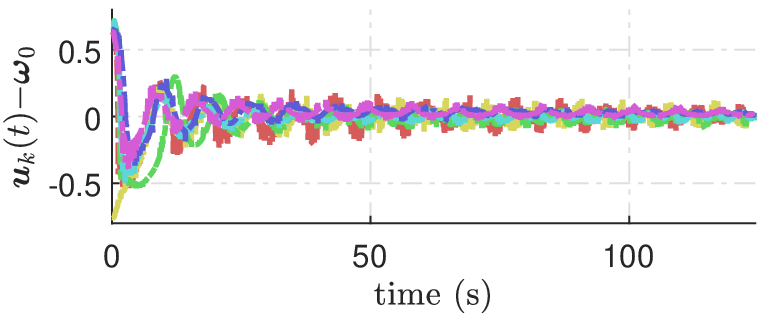}
     \caption{The cost inputs of Case \#\,3}
     \label{fig:exp_input_3}
     \end{subfigure}
     \caption{Experimental results with top-to-bottom screenshots: (a)-(c) are the Voronoi partitions of the ultimate LOCs, with the virtual centers (dark small circles), the Voronoi centroids (shadow small circles), and the orbits of the robots (thin large circles), (d)-(f) are the coverage costs, and (g)-(i) are the control inputs.}
     \label{fig:exp_result}
\end{figure*}

The results are shown in Fig.~\ref{fig:exp_result}. We only show the ultimate virtual centers, Voronoi centroids, and circular orbits in Fig.~\ref{fig:exp_image_1}, Fig.~\ref{fig:exp_image_2}, and Fig.~\ref{fig:exp_image_3} for brevity. 
The backgrounds are filled with top-to-down screenshots. All virtual centers coincide with their Voronoi centroids and all robots orbit around these coinciding positions, indicating the achievement of optimal coverage. 
Fig.~\ref{fig:exp_cost_1}, Fig.~\ref{fig:exp_cost_2}, and Fig.~\ref{fig:exp_cost_3} show that the coverage costs monotonously decay to zero for all cases. The control inputs shown in Fig.~\ref{fig:exp_input_1}, Fig.~\ref{fig:exp_input_2}, and Fig.~\ref{fig:exp_input_1} are strictly limited by $|u_k(t) \!-\! \omega_k| \!<\! \gamma_k \omega_k \!=\! 0.8$, for $k=1,2,\cdots,6$, implying the satisfaction of the input saturation constraints. This validates the applicability of the proposed control method on real robot platforms. Comparing Fig.~\ref{fig:exp_result} and Fig.~\ref{fig:init}, it is noticed that the simulation and the experimental studies have different ultimate Voronoi partitions and LOCs, even under the same initial conditions and with the same control parameters. This implies that uncertainties such as the network delay, the friction forces, and the system noise may affect which LOC is ultimately reached, even though the proposed controller is sufficiently robust to ensure the convergence to a certain LOC.
A video of the experiment is published at \href{https://youtu.be/NAvVDMRWqN8}{https://youtu.be/NAvVDMRWqN8}.


\section{Discussion}

Compared to the general coordination control of a multi-agent system with complex agent dynamics, additional challenges for optimal coverage control of a multi-CSUR system include the non-convex coverage metric function, the state- and input-dependent constraints, and the distributed realization. This paper provides the first feasible solution that solves all these issues. 
Firstly, the novel coverage cost allows us to overcome the limitation of the conventional gradient-based coverage controller for multi-SIR systems when applied to multi-CSUR systems. Secondly, a novel measurement-based method allows us to design a distributed controller. Thirdly, the Sigmoid function ensures smooth control inputs with saturations. Although not elaborating on all possible conditions in the experimental studies, the effectiveness of the proposed method has been validated for various initial conditions, control parameters, and system scales, even with uncertainties. Its critical advantage in resolving the feasibility issue is addressed compared to the conventional control method. Theoretical proofs of invariance and stability ensure it can be generalized to a wider range of robotic systems with similar dynamic models. Thus, the efficacy and applicability of the proposed method can be confirmed in a generic sense.

The proposed controller is promising to be directly applied to real fUAVs and extended to large-scale multi-robot systems, despite the usage of wheeled robots instead of real fUAVs in the experimental study, due to the lack of hardware devices and experimental spaces. Implementing the controller on fUAVs is straightforward since both robots have a similar control mechanism, i.e., driving at constant speeds with desired steering angles as control inputs. 
The main issue is handling the modeling uncertainties caused by aerodynamic factors, which can be easily solved using existing robust motor-level controllers~\cite{zhang2022tracking}. Additionally, the potential extension of the proposed method to large-scale systems is validated in the simulation study in Sec.~\ref{sec:large}. Even though we could not show it in real experiments due to the shortage of robots, we can confirm the feasibility of this extension considering the analysis of the computational complexity in Sec.~\ref{sec:dci}. The main concern is uncertainty-handling which can be solved by the existing robust control methods mentioned above.

The proposed method only applies to convex regions since Voronoi partitioning is nontrivial in nonconvex regions, such as the ones with non-standard or irregular shapes. Instead of a unique limitation of our work, optimal coverage in a nonconvex region is a common challenge for all gradient-based coverage controllers~\cite{schwager2009optimal}. Resolving this issue is beyond the scope of this paper, although we can give some hints on possible solutions. A promising approach may be splitting a nonconvex region into several convex regions and solving the coverage problems individually, for which splitting methods for nonconvex optimization problems~\cite{li2015global} may be used.

\section{Conclusion} \label{sec:con}
We propose a novel distributed optimal coverage controller for a multi-agent system with complex agent models. We have comprehensively used various theoretical tools including BLF, Lyapunov asymptotic stability, and the invariance theory to address the efficacy of this method in theory. We have also validated its practical applicability and advantage over the conventional method in simulation and experiment studies. This promising work may inspire the coverage control of a generic multi-agent system with complex agents. On the other hand, the method is limited to convex regions and lacks collision handling, highlighting the potential future work.
	
\appendices

\section*{Acknowledgment}

Q. Liu, Z. Zhang, and N. Le have shared equal contribution to this paper. Liu has led the project, provided the related work review, proposed the problem formulation and general solutions, and specified the structure of this paper. Zhang has been responsible for the main technical results and the major writing of this paper, including the preliminaries, the proposed coverage metric function, the stability and invariance analysis, and the distributed algorithm. He also provided figures and result analysis of the case studies. Le has proposed the concept of using BLF to address the state constraints and scaling gain to handle the input constraint. His bachelor thesis was an important foundation of this work. He has also been devoted to the simulation and experiments. The code and data for all simulation and experiment studies are published in \href{https://zenodo.org/record/7600131}{https://zenodo.org/record/7600131}.

\ifCLASSOPTIONcaptionsoff
\newpage
\fi




%
\medskip
\bibliographystyle{IEEEtran}
\bibliography{IEEEabrv, Reference.bib}

\end{document}